\documentclass[12pt]{article}
\usepackage{graphicx,bm,amsmath,amssymb,latexsym,psfrag,epsfig,subfigure,euscript, multirow,color}

\allowdisplaybreaks

\def\nn{ {\red{n}} }
\def\ni{ {\red{n_i}} }
\def\nJ{ {\red{n_J}} }
\def\nb{ {\red{{\bar n}}}}
\def\nbi{ {\red{{\bar n}_i}}}
\def\nslash{\rlap{\hspace{0.02cm}/}{\nn}}

\def\nslashi{\rlap{\hspace{0.02cm}/}{\ni}}
\def\nslashJ{\rlap{\hspace{0.02cm}/}{\nJ}}
\def\nbslashi{\rlap{\hspace{0.02cm}/}{\nbi}}

\newcommand{\non}{\nonumber}

\def\rd{\mathrm{d}}

\def\ecm{E_{\mathrm{CM}}}
\def\qq{ {q \bar{q}} }

\def\MX{M_X}
\def\Mx{m_X}
\def\MXt{M_X^2}
\def\Mxt{m_X^2}
\def\px{P_X}
\def\cdt{ \hspace{-0.1em} \cdot \hspace{-0.1em} }
\def\cO{ {\mathcal O}}

\def\x{x}
\def\xx{\xi}

\def\vb{\bar{v}}

\def\rd{\mathrm{d}}
\def\rc{\mathrm{c}}
\def\ptmax{p_T^{\mathrm{max}}}

\definecolor{darkred}{rgb}{0.6,0.0,0.0}
\definecolor{darkblue}{rgb}{0.0,0.0,0.5}
\definecolor{darkgreen}{rgb}{0.0,0.5,0.0}
\definecolor{brown}{rgb}{0.0,0.0,0.0}
\newcommand{\red}{\color{darkred}}
\newcommand{\blue}{\color{darkblue}}

\newcommand{\sighatab}{\frac{\rd^2 \hat{\sigma}_{a b}}{\rd w \rd v}}
\newcommand{\sighat}{\frac{\rd^2 \hat{\sigma}}{\rd w \rd v}}

\begin{document}

\begin{titlepage}

\begin{flushright}
\end{flushright}

\vspace{0.2cm}
\begin{center}
\Large\bf
Direct photon production with effective field theory
\end{center}

\vspace{0.2cm}
\begin{center}
{\sc Thomas Becher$^a$ and Matthew D. Schwartz$^{b}$}\\
\vspace{0.4cm}
{\sl $^a$\,Institute for Theoretical Physics \\
University of Bern\\
Sidlerstrasse 5, 3012 Bern, Switzerland \\[0.3cm]
$^b$\,Department of Physics\\
Harvard University\\
Cambridge, MA 02138, U.S.A.}
\end{center}

\vspace{0.2cm}
\begin{abstract}\vspace{0.2cm}
\noindent 
The production of hard photons in hadronic collisions is studied using 
Soft-Collinear Effective Theory (SCET). This is the first application
of SCET to a physical, observable cross section involving energetic partons in more than two directions. A factorization formula is derived which involves  a non-trivial interplay of the angular dependence in the hard and soft functions, both quark and gluon jet functions, and multiple partonic channels. The relevant hard, jet and soft functions are computed to one loop and their anomalous dimensions are determined to three loops. The final resummed inclusive direct photon distribution is valid to next-to-next-to-leading logarithmic order (NNLL), one order beyond previous work.  The result is improved by including non-logarithmic terms and photon isolation cuts through matching, and compared to Tevatron data and to fixed order results at the Tevatron and the LHC. The resummed cross section has a significantly smaller theoretical uncertainty than the next-to-leading fixed-order result, particularly at high transverse momentum.
\end{abstract}
\vfil

\end{titlepage}

\section{Introduction}
The production of hard photons in high energy collisions is one of the most fundamental processes 
to be observed at any hadronic collider. Photons which are produced from the underlying partonic interaction
are called direct, or prompt photons. They 
can probe the structure of the proton at very small distance scales, and are therefore sensitive both to details of the standard model and to 
possible new physics scenarios. 
At lowest order, there are two partonic processes which can produce direct photons:
the annihilation channel $\qq \to \gamma g$ and the Compton channel $q g \to q \gamma$. The Compton
channel is particularly important as it gives direct access to the gluon parton-distribution function (PDF).

On the experimental side, the photon spectrum can be measured with great precision. However, it is in general not possible to distinguish whether the photons are direct, that is they have come from the underlying hard interaction, or if have been produced
from secondary fragmentation, such as $\pi^0$ decay. This ambiguity is lessened somewhat for very high
energy photons, which are relatively unlikely to have been produced from fragmentation. Moreover, demanding
a mild isolation criterion on the photon, for example, that there be less than 2 GeV of hadronic energy 
in some reasonable surrounding area, makes the high $p_T$ photon spectrum a fairly clean probe of the underlying interaction.

On the theoretical side, the direct photon spectrum has been approached both in fixed order perturbation theory and
with soft gluon resummation. The cross section is known for both the polarized and unpolarized case at next-to-leading
order (NLO)~\cite{Aurenche:1983ws,Aurenche:1987fs,Gordon:1993qc}, in the fully inclusive case including analytic integration over the real emission contribution. The Monte Carlo program {\sc jetphox}  \cite{Catani:2002ny} implements the NLO result numerically, 
as well as the contamination from fragmentation, and allows the user to specify an isolation criterion. 

Near the partonic threshold, where the transverse momentum  $p_T$ of
the photon is close to half of the partonic center of mass energy
$p_T\lesssim \sqrt{\hat{s}}/2$, the invariant mass of the recoiling
hadronic system becomes small and the partonic cross section involves
large logarithms.
These logarithmic terms, which arise from soft
and collinear radiation, often amount to the bulk of hadronic cross
sections. To improve predictions, 
threshold contributions can be resummed to all orders in perturbation
theory. For direct photon production, this has been done to next-to-leading logarithmic order (NLL)~\cite{Laenen:1998qw,Catani:1998tm,Catani:1999hs,Kidonakis:1999hq}
and a phenomenological comparison to data from E-706 and UA-6 has
been made~\cite{Catani:1999hs,Kidonakis:2003bh}. The resummation
effects are important at large $p_T$, and therefore must be understood
to improve the precision of theoretical predictions for the direct
photon $p_T$ spectrum and related observables.

The approach we take to resummation in direct photon production is based on the use of effective field theory techniques. Effective field theories are powerful tools for separating physics associated with different scales and resumming large logarithms of ratios of those scales through the renormalization 
group. In this paper, we apply Soft-Collinear Effective Theory (SCET) \cite{Bauer:2000yr,Bauer:2001yt,Beneke:2002ph} to direct photon production. 
The effective theory was originally developed to analyze $B$-decays, but its promise for collider physics was envisioned early on \cite{Bauer:2002nz}.
The collider applications of SCET have included deep-inelastic
scattering (DIS)~\cite{Manohar:2003vb,Becher:2006nr,Becher:2006mr,Chen:2006vd},
Drell-Yan~\cite{Idilbi:2005ky, Becher:2007ty}, Higgs
production~\cite{Idilbi:2005ni, Ahrens:2008qu, Ahrens:2008nc},
$t\bar{t}$ production and event shapes in  $e^+e^-$
collisions~\cite{Lee:2006nr,Fleming:2007qr,Schwartz:2007ib,Bauer:2008dt,Becher:2008cf,Hornig:2009vb}, and electroweak Sudakov resummation \cite{Chiu:2007dg,Chiu:2008vv,Chiu:2009mg}.
The main result of much of this work has been to improve our understanding of the effective theory description of QCD, but in some cases it has led to qualitatively new phenomenology. For example, in~\cite{Becher:2008cf} 
the N$^3$LL resummation of the thrust distribution
was performed, leading to one of the best measurements of $\alpha_s$ and a strong model-independent bound on the gluino mass~\cite{Kaplan:2008pt}. However,
there is still much work to be done in demonstrating the power of this effective field theory at hadron colliders, and the present paper is a step in that direction.

All of the previous collider applications of SCET have involved processes with only two directions of large energy flow: in DIS, these are the proton and outgoing jet,
in Drell-Yan, they are the incoming hadrons, and in $e^+e^-$ they are the outgoing jets. 
Formal expressions involving several directions of large energy flow are 
straightforward to write down~\cite{Bauer:2006qp,Bauer:2008jx}, and the theory with multiple collinear fields has been used to perform electroweak Sudakov resummation of partonic amplitudes~\cite{Chiu:2008vv,Chiu:2009mg}, and to derive constraints on the structure of infrared singularities of gauge theory amplitudes~\cite{Becher:2009cu,Becher:2009qa,Becher:2009kw}. However, so far it has not been applied directly to a physical process, and it is an important step to derive and check a factorization theorem for a cross section involving three directions. The simplest such process at a hadron collider is direct photon production, which explains much of the motivation for the current work.

We begin in Section \ref{sec:xsec} with an overview of direct photon production, including a review of the relevant kinematics and a physical discussion of the factorization theorem. In Section~\ref{sec:EFT}, the factorization theorem is derived with SCET. 
In order to achieve resummation at the next-to-next-to-logarithmic order, which is
one order beyond previous results, we need the one-loop expressions for the hard, jet, and soft functions appearing in the factorization theorem. These are calculated
in Section~\ref{sec:HJS}. 
The relevant soft function, as defined through the factorization theorem, depends on radiation away from the direction of the outgoing jet, and we calculate it to one loop.
The gluon jet function, necessary for the annihilation channel, is also calculated at one loop. Using renormalization group (RG) invariance of the cross section, known results for some of the anomalous dimensions, and Casimir scaling of the soft function, we manage to extract the anomalous dimensions of the relevant hard, jet and soft functions to three loops. After solving the relevant RG equations, we combine these ingredients together into a closed analytical formula for the resummed direct photon distribution. In Section~\ref{sec:checks}, we show that the renormalization scale independence of the cross section implies a non-trivial cancellation among angular dependent parts associated with different scales. Section \ref{sec:scalechoice} discusses the scale choices and matching procedure. Finally,
in Section~\ref{sec:results}, we evaluate our formula numerically, comparing to Tevatron data and making predictions for the LHC.

\section{Direct photon cross section \label{sec:xsec}}
In this section, we establish some notation for kinematics of direct photon production. 
Then we review the differential cross section
in fixed-order QCD and discuss heuristically the factorization formula which we derive with SCET in Section \ref{sec:EFT}.

\subsection{Kinematics}
Let the incoming hadron
momenta be $P_1^{\mu}$ and $P_2^{\mu}$ and the photon momentum be
$p_{\gamma}^{\mu}$. We are interested in photon production at high $p_T\equiv p_T^\gamma$.
Our results will be most accurate when $p_T$ is near the {\it{machine threshold}} limit,
\begin{equation}
p_T\sim \ptmax = \frac{\ecm}{2 \cosh y}\, ,
\end{equation}
where $\ecm = \sqrt{(P_1 + P_2)^2}$ is the center of mass energy of the collision and $y$ is the photon's rapidity.
$p_T^{\mathrm{max}}$ is the maximum $p_T$ the photon can possibly have for a given $y$.
Of course, the phenomenology of direct photon production is dominated by much
smaller transverse momenta, but the factorization theorem will only have exact perturbative
scale independence for $p_T \sim p_T^{\mathrm{max}}$, and for its derivation we will expand around this threshold.

Near threshold, the recoiling radiation $X$ must have $P_T^X \sim p_T^\gamma \sim \ptmax$, which is only possible if the mass of the recoiling radiation is close to zero. By momentum conservation,
\begin{equation}
P_1^{\mu} + P_2^{\mu} = p_{\gamma}^{\mu} + \px^\mu\,,
\end{equation}
and the threshold implies $E_X \sim \ptmax \gg \sqrt{\px^2}$.
Near this limit, the recoiling radiation can be characterized as a jet of collinear particles with momentum $p_J^{\mu}$ accompanied by soft radiation with momentum $k^{\mu}$, $\px^\mu= p_J^{\mu}+k^{\mu}$. As we will  discuss in the next section, SCET provides a 
field-theoretic description of the associated collinear and soft partons and their interactions.

At leading order, there are two channels for direct photon production: the
Compton process $q g \rightarrow q \gamma$ and the annihilation process $\qq \rightarrow g \gamma$. In either case, let the incoming partons have
momenta $p_1^{\mu} = \x_1 P_1^{\mu}$ and $p_2^{\mu} = \x_2 P_2^{\mu}$. 
The hadronic and partonic Mandelstam variables are
\begin{equation}
s = (P_1 + P_2)^2, \hspace{1em} t = (P_1 - p_{\gamma})^2, \hspace{1em} u =
(P_2 - p_{\gamma})^2 \,,
\end{equation}
and
\begin{equation}
\hat{s} = (p_1 + p_2)^2 = \x_1 \x_2 s, \hspace{1em} \hat{t} = (p_1 -
p_{\gamma})^2 = \x_1 t, \hspace{1em} \hat{u} = (p_2 - p_{\gamma})^2 = \x_2 u \, .
\end{equation}
For direct photon production, it is conventional to work not in terms of the Mandelstam variables, but in terms of dimensionless ratios of them.
\begin{equation}
v = 1 + \frac{\hat{t}}{\hat{s}}
\, , 
\hspace{1em}
w = -  \frac{\hat{u}}{\hat{s} + \hat{t}}\,.
\end{equation}
We will also use the shorthand
\begin{equation}
\vb \equiv 1 - v
\end{equation}
for compact notation. It follows that
\begin{equation}
\hat{s} = \frac{1}{w} \frac{p_T^2}{v \vb},
\hspace{1em} 
\hat{t} = -  \frac{1}{w} \frac{p_T^2}{v},
\hspace{1em} 
\hat{u} = -  \frac{p_T^2}{\vb},
\end{equation}
\begin{equation}
\x_1 = \frac{1}{w} \frac{p_T}{\ecm v} e^y \,,
\hspace{1em} 
\x_2 = \frac{p_T}{\ecm \vb} e^{- y} \, .
\end{equation}
At the hadron level, the event is characterized by two quantities, $p_T$ and $y$. At the parton level, 
it takes four, for example,  $\{p_T, y, \x_1,\x_2\}$, or  $\{p_T, y, v, w\}$.

To understand the thresholds, it is helpful to define the hadronic invariant mass
\begin{equation}
\MXt = \px^2= (P_1 + P_2 - p_{\gamma})^2=s + t + u \,
\end{equation} 
and the partonic invariant mass
\begin{equation}
\Mxt = (p_1 + p_2 - p_{\gamma})^2 = \hat{s} + \hat{t} + \hat{u} \,.
\end{equation}
The partonic invariant mass, $\Mx$, includes only the partons involved in the hard scattering process, while the hadronic mass, $\MX$, includes also the proton remnants. 
Note that while $\MX$ is observable, $\Mx$ must be integrated over in any measurable quantity.
In the literature (e.g. in \cite{Laenen:1998qw}), the above two quantities are sometimes denoted by $S_4=\MXt$ and $s_4=\Mxt$.
These quantities represent the mass of everything in the final state except the photon, at the hadron and parton levels respectively.
At leading order in perturbation theory, where the partonic final state consists of a single parton, $w=1$ and $\Mx = 0$ exactly.

In terms of $p_T, y, v$ and $w$ the threshold variables read
\begin{equation}
\MXt = \ecm^2 - 2 p_T \ecm \cosh y = \ecm^2\left(1-p_T/\ptmax \right)
\end{equation}
and
\begin{equation}
\Mxt = \frac{p_T^2}{\vb} \frac{1 - w}{w}\,.
\end{equation}
In terms of $\Mx,\x_1,\x_2$ and $v$,
\begin{equation}
\MXt = \frac{\Mxt}{\x_2} + \ecm^2 \left[ (1-\x_1) v + (1-\x_2) \vb \right]\,.
\end{equation} 
Since partonic configurations are specified by four variables, surfaces of constant $\MX(p_T,y)$ are three dimensional. 
That is, there are three independent ways we can have the a small deviation from $\MX=0$. 
It is natural to take the independent variations to be $\x_1$, $\x_2$ and $\Mxt$ because
setting $\x_1=\x_2=1$ and $\Mx=0$ forces $\MX=0$ exactly. 
The fourth variable, $v$, can be thought of as moving us along the surface of constant $\MX$.
As $\MX \to 0$, $v\vb\to p_T^2/\ecm^2$. Then, to first order in $1-\x_1$, $1-\x_2$ and $\Mxt$,  
\begin{align} \label{MXmx}
\MXt  = \Mxt + \frac{p_T^2}{v \bar{v}} \left[ (1-\x_1)v + (1-\x_2)\vb \right] + \dots\,.
\end{align}
This is equivalent to the threshold expansion in~\cite{Laenen:1998qw}. This form for $\MX$ will be convenient for checking the SCET factorization theorem in Section~\ref{sec:checks}.

Note that the limit $\MX \rightarrow 0$ automatically enforces that the reaction takes place at the threshold $\x_1 \rightarrow 1$, $\x_2 \rightarrow 1$, where the leading partons carry almost all of the proton momentum.  In contrast, taking $\Mx \rightarrow 0$ does not force $\x_1\rightarrow 1$ or $\x_2 \rightarrow 1$. 
At the partonic level, the factorization theorem will resum logs of $\Mx$, 
which appear as $\alpha_s^n \ln
^m (1-w)$. 
It will also resum logs from the evolution of the parton distribution functions, of the form $\alpha_s^n\ln
^m(1-x_i)$, which
are only relevant near the machine threshold.

\subsection{Differential cross sections \label{sec:ds}}
Using $v$ and $w$, the cross section can be written in the form~\cite{Gordon:1993qc}
\begin{equation}
\frac{\rd^2 \sigma}{\rd y \rd p_T} 
=  \frac{2}{p_T}  \sum_{ab}
\int^{1 -  \frac{p_T}{\ecm} e^{- y}}_{\frac{p_T}{\ecm} e^y} \rd v 
\int_{\frac{p_T}{\ecm} \frac{1}{v} e^y}^1 \rd w 
\left[ \x_1 f_{a/N_1} (\x_1, \mu) \right] \left[ \x_2 f_{b/N_2} (\x_2, \mu) \right]
\sighatab \, , \label{csec}
\end{equation}
where the sum is over the different partonic channels.

At leading order the mass of the final state is zero, $w=1$, and
\begin{equation}
\sighatab= \widetilde\sigma_{ab}(v) \delta(m_X^2) =\frac{\vb}{p_T^2} \, \widetilde\sigma_{ab}(v) \,\delta(1-w)
\end{equation}
where 
\begin{align} \label{sigdef}
\widetilde\sigma_{q \bar q}(v) &= \pi \alpha_{\mathrm em} e_q^2 \alpha_s (\mu) \frac{2 C_F}{N_c}  \left( v^2 + \vb^2 \right)\frac{1}{\vb}\,, \\
\widetilde\sigma_{q g} (v) &= \pi \alpha_{\mathrm em} e_q^2 \alpha_s (\mu) \frac{1}{N_c}\non
\left( 1 + \vb^2 \right) \frac{v}{\vb}\,.
\end{align}
Here, $e_q$ are the charges of the quarks and $N_c$ is the number of colors.

At next-to-leading order (NLO), the partonic cross section acquires $w$ dependence. It has the form 
(leaving the partonic indices $ab$ implicit)
\begin{multline}
\sighat
=  \frac{\vb}{p_T^2} \,\widetilde\sigma(v) \left\{ \delta(1-w) + \alpha_s(\mu)\left[ \delta(1-w) h_1(v) + \left[ \frac{1}{1 - w} \right]_+  h_2(v) \right. \right. \\ \left.\left. + \left[\frac{\ln(1 - w)}{1 - w} \right]_+ h_3(v)+h_4(v,w) \right]\right\}
\, .
\end{multline}
The plus distributions indicate the singular behavior at NLO as $w\to1$, that is, as the kinematic threshold is approached. An N$^n$LO computation
would lead to higher-order plus distributions, up to $\left[\frac{\ln
^{2n-1}(1-w)}{1-w}\right]_+$. Keep in mind that there is implicit, non-singular, $w$ dependence in the PDF $f_a(\x_1,\mu)$ as well.
These functions $h_i$ can be found in \cite{Gordon:1993qc}. The singular ones, $h_1,h_2$ and $h_3$, as well as the singular coefficients
at NNLO are listed in Appendix \ref{app:nlo}.

The result from effective theory, which we derive in Section \ref{sec:EFT}, has the form\footnote{
The $w$ prefactor in this equation is a convention, 
but it follows from the $\hat{s}^{-1}$ dependence of the partonic cross section and conveniently
cancels the $w$-dependence of $x_1$ in Eq.~(\ref{csec}).}
\begin{equation}\label{sigmafact}
\sighat
=w\, \widetilde\sigma(v)\, H (p_T, v ,\mu) \int \rd k\, J (\Mxt - (2 E_J) k,\mu)\, S (k,\mu) \, .
\end{equation}
Here, $H$, $J$, and $S$ are the hard, jet and soft functions, respectively and $E_J$ is the energy of the jet.
The functions $H$, $J$, and $S$ are different in the two partonic channels. The hard function comes from matching SCET to QCD. It depends only on $v$, since $w=1$ at the hard scale $\mu_h\sim p_T$. The jet function comes from integrating out collinear modes. It depends on $w$ through $\Mxt=\frac{p_T^2}{1-v}\frac{1-w}{w}$. In particular, at leading order $J(p^2)=\delta(p^2)$, which reproduces the $\delta(1-w)$ dependence of the LO
cross section. Expanding the hard, jet and soft functions to order $\alpha_s$ will reproduce the other terms in the NLO cross section, up to terms which are regular as $w\to1$ and correspond to power corrections in the effective theory. Expanding
to order $\alpha_s^2$ produces all the singular terms at NNLO, which is a new result. 

The scale $\mu$ in Eq.~(\ref{sigmafact}) should be identified with the factorization scale since it determines where the PDFs are evaluated.  Since the physical scales entering $H$, $J$, and $S$ are quite different, any choice of $\mu$ will lead to large perturbative logarithms. To resum these, we will solve the RG equations for the three functions in Section \ref{sec:HJS} and evolve each of them from a matching scale to the reference scale $\mu$ at which the different contributions are combined. For the matching scale for the the hard function, we choose $\mu_h = p_T$.  The choice of the matching scales for the jet function and soft functions is less obvious, and will be discussed in Section~\ref{sec:scalechoice}.

The form of the SCET factorization theorem, Eq.~(\ref{sigmafact}), can be understood from simple physical arguments. 
The recoiling radiation $X$ in a high-$p_T$ direct photon event is almost massless. That is, $\px^2 \ll E_X$. Thus, this radiation
consists of particles forming a jet, with momentum $p_J^\mu$ and additional soft radiation $k^\mu$.
Then,
\begin{equation}
\Mxt =p_X^2= (p_J + k)^2 = m^2 + 2 k \cdt p_J \label{mxmk}
\end{equation}
up to terms of order $k^2 \ll m^2$, where $m^2=p_J^2$ is the mass of the jet.  The precise allocation of the final state
particles into the jet or the soft sector is not well defined, but the ambiguities give only power suppressed corrections to this relation.

Since the jet is lightlike at leading order, its momentum can be written as $p_J^{\mu} \sim E_J {\red n_J^{\mu}}$, where ${\red n^{\mu}_J} = (1,{\red \vec{n}_J})$ is a lightlike vector. This is true up to power corrections, because ${\red\bar{n}_J}\cdt p^J \gg p^J_{\perp} \gg {\red n_J}\cdt p^J$. Thus, the amplitude for producing a configuration with a particular
value of $\Mxt$ will be proportional to
\begin{equation}
J (m^2) = J \left( \Mxt - (2E_J) ({\red n_J} \cdt k)\right) \,.
\end{equation}
This explains the $E_J$ dependence in Eq.~(\ref{sigmafact}). Also, we see that 
the only component of the soft radiation which is relevant to
threshold resummation is the ${\red n_J}\cdt k$ component, that is, the piece backwards to the direction of the jet. This projection
$k\equiv ( {\red n_J} \cdt k)$ also appears in the soft function of Eq.~(\ref{sigmafact}). 
Our derivation of the factorization theorem in the next section will provide us with operator expressions for the jet and soft functions appearing in the factorization theorem. We will compute these functions to one loop in Section \ref{sec:HJS}.

\section{Derivation of the factorization theorem\label{sec:EFT}}
We will split up the derivation into two parts. First, we will summarize some results from SCET about operators and scaling relations. Then we will apply the
effective theory to the process of direct photon production and derive the factorization theorem for the differential cross section.

\subsection{Soft-Collinear Effective Theory}
SCET provides an expansion in the limit of large energies and small invariant masses.
For a jet with momentum $p_J^\mu$, we expand in
$\varepsilon = m_J/E_J$.
For direct photon production, there are three
high-energy scales, the energy of the two incoming partons and the energy of the hadronic final state $X$.
Correspondingly, we introduce three different sets of collinear fields associated with the directions of large energy flow and one set of soft fields, 
which mediate interactions among the different collinear directions. 

The components of momenta $p_{\rc i}^\mu$ of collinear fields associated with the $i$-th direction, whether
quark ($\psi_i$), antiquark ($\bar\psi_i$) or gluon ($A_{ i}^\mu$), scale as
\begin{equation}\label{collscaling}
\mbox{$i$-collinear:} \quad
(\nbi \cdt p_{\rc i},\, \ni \cdt p_{\rc i},\, \, p_{\rc i  \perp}^\mu) 
\sim (1, \varepsilon^2, \varepsilon) \,,
\end{equation}
where the vectors ${\red n_i^\mu}$ are light-like reference vectors along the $i$-th direction. For each light-cone vector ${\red n_i^\mu}$, 
we also introduce a conjugate light-cone vector, ${\red \bar{n}_i^\mu}$, such that $\nbi\cdt\ni = 2$. The field components of a collinear gluon scale in exactly the same way as its momentum, due to gauge invariance.
Two components of collinear Dirac spinors are suppressed and can be integrated out, after which the spinors fulfill the constraint $\nslash_{\red{i}}\, \chi_i = 0$. 
Then, collinear fermion fields scale as $\varepsilon$.
All components of soft momenta, with respect to any of the jet directions, are small
\begin{equation}\label{softscaling}
\mbox{soft}: \quad p_{\mathrm{s}}^\mu \sim \varepsilon^2\, .
\end{equation}
Thus, soft fields can interact with any of the collinear fields without changing their scaling. Soft gluon and soft quark fields scale as ${A}_{s}^\mu \sim \varepsilon^2$ and $\psi_s \sim \varepsilon^3$,
respectively.

To construct operators in the effective theory, it is convenient to work with the jet fields $\chi_i$ and ${\cal A}_{i\perp}^\mu$ \cite{Bauer:2001yt,Hill:2002vw}. They describe the propagation of energetic partons in the $i$-th direction, but do not correspond to any experimental definition of jet, such as cone or $k_T$-jet.  Explicitly, the jet fields are related to free collinear quark and gluon fields by the addition of Wilson lines
\begin{equation}\label{eq:collfields}
\chi_i(x) = W_i^\dagger(x)\,\frac{\nslashi \nbslashi} {4} \,     \psi_i(x) \,, \qquad 
{\cal A}_\perp^\mu(x) = W_i^\dagger(x)\,[iD_\perp^\mu W_i(x)] \,.
\end{equation}
These $i$-collinear Wilson lines
\begin{equation}\label{eq:Whc}
W_{i}(x) = {\rm\bf P}\,\exp\left(ig\int_{-\infty}^0\!ds\, \nbi\cdt A_i(x+s\nbi) \right)
\end{equation}
ensure that fields are invariant under collinear gauge transformations in each sector \cite{Bauer:2000yr,Bauer:2001yt}. The symbol ${\bf{P}}$ indicates path ordering, and the 
conjugate Wilson line $W_i^\dagger$ is defined with the opposite ordering prescription.

At leading power only the $\ni\cdt A_s$ component of the soft field can interact with the collinear fields in the $i$-th direction,
since all other components are power suppressed compared to components  of the collinear gluon field. As a consequence, the leading-power soft-collinear interactions are Eikonal and the soft dynamics can be removed from the collinear Lagrangian to all orders in perturbation theory through field redefinitions. 
For example, the interaction of soft gluons with collinear fermions in the SCET Lagrangian has the form
\begin{equation}\label{Lcs}
{\cal L}_{c_i+s} = \bar\chi_i(x)\,\frac{ \nbslashi}{2}\,
\ni\cdt A_s(x_{-})\,\chi_i(x) \,.
\end{equation}
where $x_{-}^\mu = ({\red \bar{n}_i}\cdt x )\frac{\red n_i^\mu}{2}$.  The peculiar $x$-dependence of the soft field will be explained below. This interaction can be represented in terms of soft Wilson lines. Redefining the quark and gluon fields as
\begin{align}\label{decouple}
\chi_i(x) &\to Y_i(x_-)\,\chi_i(x) \,, \\
\bar\chi_i(x) &\to Y_i^\dagger(x_-)\bar\chi_i(x) \,, \\
{\cal A}_{i\perp}^\mu(x)    &\to Y_i(x_-)\,{\cal A}_{i\perp}^\mu(x) Y_i^\dagger(x_-) \,,
\end{align}
where 
\begin{equation}\label{softwilson}
Y_i(x) = 
{\bf P} \exp\left( ig \int_{-\infty}^0\,\rd t\, \ni\cdt A_s^a(x+t \ni)\,t^a \right) ,
\end{equation}
eliminates the interaction ${\cal L}_{c_i+s}$ and other pure-gluon terms. 
After this decoupling transformation, soft interactions manifest themselves only through Wilson lines in the operators~\cite{Bauer:2001yt}.

Let us now explain why the collinear fields in soft-collinear interactions, such as in Eq.~(\ref{Lcs}), 
are evaluated at $x$ and the soft fields at $x_{-}$. 
First, recall that a collinear sector alone should be equivalent to full QCD (it can be derived as QCD in a boosted frame). Therefore,
no information must be lost in the derivative (or "multipole") expansion of a collinear field \cite{Beneke:2002ph}
\begin{equation}
\psi_i(x) = 
\left[1+\frac{1}{2} ({\red \bar{n}_i}\cdt x )\partial_{\red n_i} + \frac{1}{2} ({\red n_i}\cdt x )\partial_{\red \bar{n}_i}+ x_\perp\cdt\partial_\perp + \cdots\right]
\psi_i(0)= \psi_i(x) +\mathcal{O}(\varepsilon)\,.
\end{equation}
Since $(\partial_{\red n_i}, \partial_{\red \bar{n}_i}, \partial_\perp)\sim (\varepsilon^2, 1, \varepsilon)$, for each of these terms to to scale like $\varepsilon^0$ the scaling of $x$ is fixed:
\begin{equation} \label{xscaling}
( {\red \bar{n}_i}\cdt x, {\red n_i}\cdt x, x_\perp) \sim ( \varepsilon^{-2}, 1, \varepsilon^{-1})\,.
\end{equation}
This is collinear scaling in position space.
Now, for a soft field interacting with a collinear field at the same point $x$, we can multipole expand as well
\begin{equation} \label{sscaling}
A_s^\mu(x) = 
\left[1+ \frac{1}{2}({\red \bar{n}_i}\cdt x )\partial_{\red n_i} + \frac{1}{2} ({\red n_i}\cdt x )\partial_{\red \bar{n}_i}+ x_\perp \cdt \partial_\perp + \cdots\right]
A_s^\mu(0)=A_s^\mu\left( ({\red \bar{n}_i}\cdt x )\frac{\red n_i^\mu}{2}\right)+\mathcal{O}(\varepsilon)\,.
\end{equation}
For these fields, since the soft momenta scale like $(\partial_{\red n_i}, \partial_{\red \bar{n}_i}, \partial_\perp)\sim (\varepsilon^2, \varepsilon^2, \varepsilon^2)$, only the terms like $\left[({\red {\bar n}_i}\cdt x )\partial_{\red n_i}\right]^k$ are unsuppressed, 
which is why $A_s^\mu(x)=A_s^\mu(x_-)$ at leading power. This is simply the position space version of Eq.~(\ref{mxmk}),
$(p_J +k)^2 = m^2+2 E_J({\red n_J}\cdt k) + \mathcal{O}(\varepsilon)$, 
and will play a crucial role in the derivation of the factorization theorem below.

\subsection{SCET for direct photon production}
To study direct photon production in the effective theory we first introduce three light-like reference vectors. 
Two vectors $\red{n_1^\mu}$ and $\red{n_2^\mu}$ are aligned with the beam
and point in the direction of the incoming hadrons with momenta $P_1^\mu$ and $P_2^\mu$. 
The third reference vector $\red{n_J^\mu}$ is along the direction of the hadronic jet,
which recoils against the hard photon.

With the field content and scaling dimensions of SCET established, the first step is to match to the full standard model. 
For direct photon production, we need to introduce operators which can reproduce the matrix element of the the vector current $J^{\nu}(x)=\bar{\psi}(x)\,\gamma^\nu\,\psi(x)$ in the full
theory. The leading operators relevant for the partonic process $\qq \to \gamma g$ are
\begin{align}\label{eq:Oqq}
{\mathcal{O}_{\qq}^{S\, \nu}}(x^\mu; t_1,t_2,t_J) &= \bar{\chi}_2(x^\mu+ t_2 {\red \bar{n}_2^\mu }  ) \,{{\cal A}_J}_{\perp}^{\nu}(x^\mu+ t_J {\red \bar{n}_J^\mu }) \, 
{\chi}_1(x^\mu+ t_1  {\red \bar{n}_1^\mu } ) \, ,  \\
{\mathcal{O}_{\qq}^{T\,\nu}}(x^\mu; t_1,t_2,t_J) &= \bar{\chi}_2(x^\mu+ t_2 {\red \bar{n}_2^\mu }  ) \,i\sigma_{\nu\rho} \,{{\cal A}_J}_{\perp}^{\rho}(x^\mu+ t_J  {\red \bar{n}_J^\mu })  \,
{\chi}_1(x^\mu+ t_1 {\red \bar{n}_1^\mu }) \,.\non
\end{align}
These two operators correspond to the two cases where the quarks have equal or opposite spin. We are interested in the unpolarized cross-section, and so will need the sum of both contributions.
In addition to $\mathcal{O}_{\qq}^{S\,\nu}$ and  $\mathcal{O}_{\qq}^{T\,\nu}$ there are $5\times 2$ more operators, which are obtained by permuting the indices on the fields in $\mathcal{O}_{\qq}^{S\,\nu}$ and  $\mathcal{O}_{\qq}^{T\,\nu}$, and which describe the other partonic channels with initial states $\bar{q} q$, $q g$, $g q$, $\bar{q} g$, and $g \bar{q}$. 
Their Wilson coefficients can all be derived from the Wilson coefficients of the $\mathcal{O}_{\qq}^{\nu}$ operators by exchanging the momenta. 
For the case of the $\mathcal{O}_{qg}^\nu$ operators, associated with the $q g \to \gamma q$ channel, the corresponding crossing relations are nontrivial, and care has to be taken to get the proper imaginary parts. There are also operators with three collinear gluon fields. It is straightforward to include them, however, the $gg\to g\gamma$ channel starts contributing only at NNLO  and these operators are only relevant for N${}^3$LL resummation. In the following, we will generically refer to all the operators relevant for direct photon production as $\cO_j^\nu$.

Note that the operators $\cO_j^\nu$ are not local.
The non-locality arises because derivatives along the directions associated with large momentum flow are not suppressed. 
The variables $t_1,t_2$ and $t_J$ on which the operators depend are the position space equivalent of the label momenta introduced in \cite{Bauer:2000yr}. The smearing in the ${\red n_i^\mu}$ direction which they induce
allows for different amounts of energy in the corresponding collinear fields.
The Wilson coefficients for the operators must also also depend on these $t_i$'s, and these variables must be
integrated over in matching to the full theory
\begin{equation} \label{jmu}
J^\nu(x^\mu) = \sum_j \int \rd t_1\, \rd t_2 \, \rd t_J \,  C_j(t_1,t_2,t_J)\, \cO_j^\nu(x^\mu, t_1,t_2,t_J) .
\end{equation}
For the calculation of the cross section, we will need the Fourier transformed Wilson coefficients 
\begin{equation}\label{fourier}
{\widetilde C}(\nb_{\red 1}\cdt P_1, \nb_{\red 2}\cdt P_2,\nb_{\red J}\cdt \px) =  \int \rd t_1\, \rd t_2 \, \rd t_J \,
e^{-i \left[t_1 (\nb_{\red 1}\cdt P_1)+t_2( \nb_{\red 2}\cdt P_2)-t_J (\nb_{\red J}\cdt \px)\right]}
C(t_1,t_2,t_J)
\end{equation}
which depend on the large component of the momenta in each of the three directions. The fact that the Wilson coefficients depend on the large light-cone components of the collinear particles is characteristic for SCET. An alternative to the position space formalism \cite{Beneke:2002ph} we are using is the label formalism \cite{Bauer:2000yr}, where the large momentum component is treated as a label on the collinear fields, similar to the heavy quark velocity in HQET. 


The starting point for the factorization theorem is a generic expression for the cross section in terms of matrix elements of the production current, summed over final hadronic states, and differential in the photon momentum
\begin{equation} \label{j0eq}
\rd\sigma = \frac{2\pi \alpha_e\, e_q^2}{\ecm^2} \frac{d^3 p_\gamma}{(2\pi)^3 2 E_\gamma} \sum_X (2\pi)^4 \delta^{(4)}(P_1+P_2-\px-p_\gamma)  \big |\langle  X \, |\,\epsilon_\nu J^{\nu}(0) \, | N_1(P_1)\, N_2(P_2) \rangle \big |^2 \,.
\end{equation}
The states $|X\rangle$ are the hadronic part of the final states allowed in the process and $\epsilon_\nu$ is the photon polarization vector.
For high-$p_T$ direct photon production, 
these states must include a hard jet and so the scaling of $\px^\mu$ is like that of the jet momentum $p_J^\mu$.

For the matching step, we integrate out the hard modes of the theory. This amounts to plugging in the representation of the electromagnetic current operator in the effective theory
and restricting the final states to soft and collinear modes. 
After matching the current $J^\nu(0)$ using Eq.~(\ref{jmu}), the collinear fields in Eq.~(\ref{j0eq}) are evaluated at positions $\chi_i(t_i {\red \nb^\mu_i})$. 
These points can be translated to 
$t_i=0$, using
\begin{equation}
\chi_i( t_i {\red \nb^\mu_i}) = e^{+i t_i (\nbi \cdot {\bm P}_i )}\, \chi_i(0) e^{-i t_i (\nbi\cdot  {\bm P}_i) }\,.
\end{equation}
Then we can have the momentum operators act on the states, where they evaluate to the large momentum associated with the given direction. Performing
the integral over the convolution variables $t_i$, as in Eq.~(\ref{fourier}), yields the Fourier transforms of the 
hard matching coefficient. Thus, we have
\begin{multline}
\rd\sigma = \frac{2\pi \alpha_e\, e_q^2}{\ecm^2} \frac{\rd^3 p_\gamma}{(2\pi)^3 2 E_\gamma}  \sum_X
\int \rd^4 x\, e^{i(P_1+P_2-\px-p_\gamma)x} \,\\
\times  \Big |  \sum_{j}
{\widetilde C}_j (\nb_{\red 1}\cdt P_1, \nb_{\red 2}\cdt P_2,\nb_{\red J}\cdt \px)
\langle X \, | \epsilon_\nu \cO_{j}^\nu(0) | N_1(P_1)\, N_2(P_2) \rangle \Big|^2 \,, \label{interm}
\end{multline}
where ${\mathcal{O}^\nu_{j}}(x)\equiv {\mathcal{O}^\nu_{j}}(x^\mu;0,0,0)$. 
Note that this has been separated into a sum over states $|X\rangle$ with only soft and collinear fields, and a sum over the operators.

We would like to get rid of the sum over states and write the above expression (\ref{interm}) as a forward matrix element. To this end, we turn the $\exp(i P_i x)$ factors to operators $\exp(i {\bf P}_i x)$, by using the states $| N_i(P_i)\rangle$ and $|X\rangle$. Then these operators act on the the three collinear fields in $\cO_j^\nu(0)$, moving the entire operator to $\cO_j^\nu(x)$. After summing over photon polarizations, this gives
\begin{multline}\label{eq:matrixelement}
\rd\sigma = \frac{2\pi \alpha_e\, e_q^2}{\ecm^2} \frac{\rd^3 p_\gamma}{(2\pi)^3 2 E_\gamma} \, 
\sum_X \sum_{j,k}
{\widetilde C}^*_j (\nb_{\red 1}\cdt P_1, \nb_{\red 2}\cdt P_2,\nb_{\red J}\cdt \px)
{\widetilde C}_k (\nb_{\red 1}\cdt P_1, \nb_{\red 2}\cdt P_2,\nb_{\red J}\cdt \px)\\
\times  \int \rd^4 x\, e^{-i (p_\gamma x)}\,
\langle  N_1(P_1)\, N_2(P_2) |\,  \cO_j^{\nu\dagger}(x)\, |X\rangle\langle X|\, \cO_k^\nu(0)  \,|  N_1(P_1)\, N_2(P_2) \rangle \,.
\end{multline}
This has the form of a two point function, where the interaction between $\cO_j^{\nu\dagger}(x)$ and $\cO_k^{\nu}(0)$ is mediated by exchange of final state particles $|X\rangle$. Since $P_X^\mu$ scales like a collinear
field in the ${\red n_J^\mu}$ direction, $x$ must scale like the conjugate position space coordinate (see Eq.~(\ref{xscaling})):
\begin{equation}
( {\red \bar{n}_J}\cdt x, {\red n_J}\cdt x, x_\perp) \sim ( \varepsilon^{-2}, 1, \varepsilon^{-1})\,.
\end{equation}
This scaling will help define the soft function below.

Next, we perform the field redefinition to remove the soft interactions from the Lagrangian.
The operators then factorize into a soft and a collinear part. For example, 
\begin{align}\label{eq:OpFac}
{\mathcal{O}_{\qq}^{S\,\nu}}
&= \left(\vphantom{Y_2^\dagger}\bar{\chi}_2  \,{\cal A}^{\nu a}_{J \perp} \chi_1\right)\left( Y_2^\dagger Y_Jt^a Y_J^\dagger Y_1\right)
=\mathcal{O}^{c\,\nu}_{\qq}\mathcal{O}^{s}_{\qq} \,,
\end{align}
where we suppress the color indices of the quark fields, which are contracted with those of the soft Wilson lines.
The collinear operators $\mathcal{O}^{c\,\nu}_j$ have the same form as the operators in Eq.~(\ref{eq:Oqq}), but are built from fields which no longer have soft interactions. The matrix elements of the collinear operators give the PDFs and the jet functions. They are sensitive to the gluon's polarization and quark spins, but diagonal in color space. 
On the other hand, the soft interactions are independent of spin, 
but inherit their color from the original process in full QCD. 
Explicitly, the soft part of the operators for the $\qq \to \gamma g$ process are given by
\begin{equation}\label{softcolor}
\left[\mathcal{O}^{s}_{\qq}(x)\right]^a_{ij} = \left[Y_2^\dagger(x) \, Y_J(x) \,t^a Y^\dagger_J(x) Y_1(x)\right]_{ij} \, ,
\end{equation}
where $i$ and $j$ are color indices.
Each of these $Y$'s is a matrix in color space, and the final soft
operator depends on the color of the quarks and the gluon. The fact
that the collinear matrix elements are color diagonal implies that in
the matrix element squared the color indices of $\mathcal{O}^{s\dagger}_j(x)$
get contracted with $\mathcal{O}^{s}_k(0)$ so that the soft function
will involve a color trace.

Because the soft and the collinear sectors no longer interact among each other, the matrix elements of the operators factorize into a product of matrix elements. Also, since $\px^\mu$ scales like $p_J^\mu$, the states $|X\rangle$ have collinear radiation in the ${\red n_J^\mu}$ direction and soft radiation, but not collinear radiation in the direction of the nucleons.
For the matrix element of $\mathcal{O}_{\qq}^{S\,\nu}$ this means
\begin{multline}
\left\langle  N_1(P_1) N_2(P_2) \left| \mathcal{O}_{\qq}^{S\, \nu \dagger}(x) \, {\mathcal{O}_{\qq}^{S}}^{\nu}(0)  \right|  N_1(P_1)\, N_2(P_2) \right\rangle \,
=\\
\left\langle  N_1(P_1) \left| \bar \chi_{1\alpha} (x) \chi_{1\beta}(0) \right|  N_1(P_1) \right\rangle \: \times \:
\left\langle  N_2(P_2) \left| \bar \chi_{2\beta} (x)  \chi_{2\alpha}(0) \right|  N_2(P_2) \right\rangle \\
\times \sum_{X_c}  
\langle  0 |  {{\cal A}_J^\nu}_{\perp}(x)  | X_c \rangle
\langle  X_c |  {{\cal A}^\nu_J}_{\perp}(0) |0 \rangle
\times
\sum_{X_s}
\langle 0 | \mathcal{O}^{s\dagger}_{\qq}(x)|X_s \rangle 
\langle  X_s|   \mathcal{O}^{s\vphantom{\dagger}}_{\qq}(0)|0 \rangle \, .
\end{multline}
Note that the Dirac indices $\alpha$ and $\beta$ are contracted among the different collinear fermions. In this factorized form, it is now obvious that the collinear matrix elements are diagonal in color space. As stated above, this implies that the color indices of the soft operator shown in Eq.~(\ref{softcolor}) are contracted between $\mathcal{O}^{s\dagger}_{\qq}(x)$ and $\mathcal{O}^{s}_{\qq}(0)$.

The matrix elements of the collinear fields associated with the jets give rise to the quark and gluon jet functions
\begin{align}\label{eq:jetdef}
\langle  0|\, \bar \chi_J^i\left( x \right) \Gamma   \chi_J^j(0) \, | 0 \rangle & 
= \delta^{ij} \,{\rm tr}\left[\frac{\nslashJ}{2} \Gamma \right] 
\int \frac{\rd^4 p}{(2\pi)^3} \theta(p^0)\, ({\red \bar{n}_J}\cdt  p ) \,J_q(p^2)\, e^{-i\, x\, p} \, , \\
\langle  0 |\, \, {{\cal A}_{J}^a}_\perp^\mu( x) {{\cal A}_{J}^b}_\perp^\nu(0)\,  | 0 \rangle & = \delta^{ab} \,
(-g_\perp^{\mu\nu})\,g_s^2 \int \frac{\rd^4 p}{(2\pi)^3}\, \theta(p^0)\,  J_g(p^2) \,e^{-i\, x\, p} \, ,
\end{align}
where $i,j$ and $a$, $b$ are the color indices of the fields. The quark jet function is the same universal object appearing in the factorization theorems for DIS~\cite{Becher:2006mr} and thrust~\cite{Fleming:2007qr,Schwartz:2007ib,Becher:2008cf}. The gluon jet function has appeared in the analysis of quarkonium production \cite{Fleming:2002rv,Fleming:2002sr,Fleming:2003gt}.

For the other matrix elements, we can simplify things by using the SCET scaling relations to project out the leading power. For the soft operator, using the multipole expansion and the scaling of $x$ 
we find that it depends only on  $x_-^\mu = ({\red \bar{n}_J}\cdt x )\frac{ {\red n_J^\mu}}{2}$ to leading power,
as in Eq.~\eqref{sscaling}. Then the soft function relevant, for example, for the partonic channel $q \bar q \to \gamma g$ reads
\begin{equation} \label{softdef}
 \langle 0 |\, {\rm Tr}\, \bar{\bm T}\! \left[( Y_1 \, Y_J t^a\,Y_J^\dagger Y_2)(x_-)\right ] \,
{\bm T}\! \left[( Y_2^\dagger \, Y_J t^a Y_J^\dagger Y_1 )(0)\right] | 0 \rangle 
= C_F N_c \,\int_0^{\infty} \rd k_+ \, e^{-i k_+ ( {\red \bar{n}_J} \cdot x)/2}\, S_{\qq}(k_+) 
\,.
\end{equation}
The prefactor on the right-hand side was chosen, such that $S_{\qq}(k_+)=\delta(k_+)$ at leading order.
Since the soft function only depends on $x_-$, its Fourier transform only depends on $k_+ = {\red n_J} \cdot k$, where $k^\mu$ is the soft radiation in $|X\rangle$. The soft function can also be written as
\begin{equation}
S_{\qq}(k_+)=  \frac{1}{C_F N_c} 
\sum_{X_s} \left |\left\langle X_s\left| 
{\bm T}\! \left[ Y_1^\dagger(0) \, Y_J(0)t^a\,Y_J^\dagger(0)Y_2(0)\right ] \,
\right| 0 \right \rangle\right |^2 (2\pi) \delta({\red n_J}\cdt p_{X_s}- k_+) \,,
\end{equation}
where the color indices of the Wilson lines need to be contracted as in Eq.~(\ref{softdef}) above.
The soft function is the amplitude squared for the emission of a set of soft partons from the three Wilson lines. The time-ordered product appears because cross sections are extracted from expectation values of time-ordered products of fields. A discussion of how expressions such as Eq.~(\ref{softdef}) arise in the path integral formulation of SCET can be found in Appendix C of Ref.~\cite{Becher:2007ty}.

For the matrix elements involving the incoming nucleons, the momenta and derivatives scale like
\begin{equation}
(\partial_{\red n_i}, \partial_{\red \bar{n}_i}, \partial_\perp)\sim ({\red n_i}\cdt P_i,{\red{ \bar{n}_i}}\cdt P_i,P_\perp)\sim  (\varepsilon'^2, 1, \varepsilon')\,,
\end{equation}
where $\varepsilon' = m_N/E_i$. We assume that the nucleon masses are negligible, $m_N \ll m_J$, so that $\varepsilon' \ll \varepsilon$. Then, including only the leading power in the multipole expansion,
$\chi_i(x)=
\chi_i\left(\ni \! \cdot\! x \frac{{\red \bar{n}^\mu_i }}{2}\right)$.
\!\!\footnote{A proper treatment of a theory with two expansion parameters $\varepsilon$ and $\varepsilon'$ would
involve messenger modes, {\it i.e.} soft modes involving the expansion parameter $\varepsilon'$. In this case, the messenger modes can be absorbed into the 
parton distribution functions. A detailed analysis of an an analogous situation has been performed for DIS in
\cite{Becher:2006mr}, and we choose
not to repeat it here.}
The expanded collinear matrix elements are the usual PDFs 
\begin{align}\label{eq:PDFdef}
\langle  N_i(P_i) | \bar \chi_i\left(
\ni \cdt x \frac{ {\red \bar{n}_i^\mu} }{2}
\right) \Gamma   \chi_i(0) \,|  N_i(P_i) \rangle & =\frac{1}{4}
\nbi \cdt P_i \, {\rm tr}\left[ \nslashi \Gamma \right] \int_{-1}^1 \rd\xx\, f_{q/N_i}(\xx)\,e^{i\,\xx\, ({\red n_i}\cdot x)(\nbi \cdot P_i)/2} \,, \non\\
\langle  N_i(P_i)\, |\,\, (-g_{\mu\nu})\,\, {\cal A}_{i \perp}^\mu\left(
\ni \cdt x \frac{ {\red \bar{n}_i^\mu} }{2}
\right) {\cal A}_{i \perp}^\nu(0) \,|  N_i(P_i) \rangle & =  
\int_{-1}^1 \frac{\rd \xx}{\xx}\, f_{g/N_i}(\xx)\,e^{i\, \xx ({\red n_i}\cdot x)\, (\nbi \cdot  P_i)/2}\,,
\end{align}
for quarks and gluons, respectively. The SCET matrix elements are identical to the PDFs defined in QCD because the collinear Lagrangian is equivalent to the original QCD Lagrangian after the decoupling. Negative values of $\xi$ correspond to the anti-particle PDF,  $f_{\bar{q}/N_i}(\xx) = \bar{f}_{q/N_i}(\xx)= - f_{q/N_i}(-\xx)$ and $f_{g/N_i}(\xx) =  \bar{f}_{g/N_i}(\xx)=-f_{g/N_i}(-\xx)$. Matrix elements which involve different fields all vanish. For example,
\begin{equation}
\langle  0\, | \bar \chi_J\left( x \right)\,  {\cal A}_{J\perp }^\nu(0)  \,|  0 \rangle = 0\,.
\end{equation}
Furthermore, because of  the traces in the collinear matrix elements, Eqs.~(\ref{eq:jetdef}) and (\ref{eq:PDFdef}), the mixed tensor-scalar matrix elements vanish as well.
Thus, only the diagonal terms $j=k$ contribute in the sum, Eq.~(\ref{eq:matrixelement}). 

Now let us combine the different ingredients. For the $\cO_{\qq}^{S\,\nu}$ operator, we get
\begin{multline}
\rd\sigma \propto
\int \rd^4 x\,
\int \frac{\rd^4 p_J}{(2\pi)^4}
\int_0^1 \rd\xx_1
\int_0^1 \rd\xx_2 
\int \rd k_+   |{\widetilde C}_{\qq}^S|^2
f_{q/N_1}(\xx_1)
f_{\bar{q}/N_2}(\xx_2)
J_g(p_J^2)
S_{\qq}(k_+) \\
\times e^{-i (p_\gamma x)}
e^{-i k_+ ({\red \bar{n}_J} \cdot x)/2}
e^{-i\, (p_J \cdot x)}
e^{i\,\xx_1 ({\red n_1}\cdot x)({\red \nb_1} \cdot P_1)/2}
e^{i\, \xx_2 ({\red n_2}\cdot x)\, ({\red \nb_2} \cdot P_2)/2} \,.
\end{multline}
The $x$ integral gives
$(2\pi)^4\delta^{(4)}(p^\mu_1+p_2^\mu-p_\gamma^\mu-p^\mu_J - k_+ \frac{{\red {\bar n}_J^\mu}}{2})$, where the parton momenta are $p_1^\mu=\xx_1 ({\red \nb_1} \cdot P_1 )\frac{{\red n_1^\mu}}{2}$ and $p_2^\mu = \xx_2 ({\red \nb_2} \cdot P_2 )\frac{{\red n_2^\mu}}{2}$. Doing the $p_J$ integral then gives
\begin{equation}
\rd\sigma \propto
\int_0^1 \rd\x_1
\int_0^1 \rd\x_2
\int \rd k
|{\widetilde C}_{\qq}|^2
f_{q/N_1}(\x_1)f_{\bar{q}/N_2}(\x_2)
J_g(\Mxt-(2E_J)k)
S_{\qq}(k) \,,
\end{equation}
where $\Mxt = (p_\gamma-p_1-p_2)^2$, $2E_J ={\red \nb_J}\cdt (p_1+p_2-p_\gamma)$,  and we have relabeled $\xx_i$ as $\x_i$ and $k_+$ as $k$.

To get to the final form of the factorization theorem, we observe that at leading order $J(p^2)=\delta(p^2)$ and $S(k)=\delta(k)$.
Thus, the sum over Wilson coefficients $\sum_j|\widetilde{C}_j^2|$,
including the factors of $2$ and such from the $\Gamma$ traces and ${\red n}\cdt{\red \bar{n}}$ factors, must reproduce
the full leading order direct photon cross section. So, we define hard functions $H_{\qq}$ and $H_{q g}$ for the two channels to be the the sum over the squares of the relevant Wilson coefficients
normalized to their values to leading order in perturbation theory. Including the appropriate Jacobian factors, 
the contribution of the annihilation channel to the cross section reads
\begin{multline}
\frac{\rd^2 \sigma_{q\bar q}}{\rd y \rd p_T} 
=  \frac{2}{p_T} 
\int^{1 -  \frac{p_T}{\ecm} e^{- y}}_{\frac{p_T}{\ecm} e^y} \rd v 
\int_{\frac{p_T}{\ecm} \frac{1}{v} e^y}^1 \rd w 
\left[ (w \x_1) f_{q/N_1} (\x_1, \mu) \right] \left[ \x_2 f_{\bar{q}/N_2} (\x_2, \mu) \right] \label{factthm1} \\
\times {\widetilde \sigma}_{\qq}(v)
H_{q\bar q}(p_T,v,\mu) \int d k J_{g} (\Mxt - (2 E_J) k,\mu) S_{q\bar q} (k,\mu)\,,
\end{multline}
and for the Compton channel
\begin{multline}
\frac{\rd^2 \sigma_{q g}}{\rd y \rd p_T} 
=  \frac{2}{p_T} 
\int^{1 -  \frac{p_T}{\ecm} e^{- y}}_{\frac{p_T}{\ecm} e^y} \rd v 
\int_{\frac{p_T}{\ecm} \frac{1}{v} e^y}^1 \rd w 
\left[ (w \x_1) f_{q/N_1} (\x_1, \mu) \right] \left[ \x_2 f_{g/N_2} (\x_2, \mu) \right] \label{factthm2} \\
\times {\widetilde \sigma}_{q g}(v)
H_{q g}(p_T,v,\mu) \int d k J_{q} (\Mxt - (2 E_J) k,\mu) S_{q g} (k,\mu)\,.
\end{multline}
The definition of $\widetilde{\sigma}$ and a simpler, more physical, discussion of this factorization formula were given in Section \ref{sec:ds}. 

To finish, let us briefly discuss the other photon-production mechanism, where the photon is produced by fragmentation. In this case, the relevant SCET operators involve four collinear fields, in the directions of the incoming hadrons and the outgoing jet as well as in the direction of the outgoing photon. The matrix element of the collinear fields in the photon direction corresponds to the fragmentation function. Since the invariant mass of the hadronic final state is small near threshold, it cannot contain any hard collinear partons in the photon direction. The outgoing collinear quark must thus fragment into the photon and a soft quark. Soft quark fields are power suppressed, which explains the smallness of the fragmentation contribution at large $p_T$.

\section{Calculation of the cross section in SCET \label{sec:HJS}}
With the factorization formula in hand, we can proceed to calculate the hard, jet and soft functions in perturbation theory.
Then we will use the RG  to run between the relevant matching scales providing the final resummed distribution.

\subsection{Hard function}
The hard functions $H_{\qq}$ and $H_{q g}$ entering the factorization
formulas, Eqs. \eqref{factthm1} and \eqref{factthm2}, are
given by the absolute value squared of the Wilson coefficients of
operators, such as ${\cal O}_j^{S \nu}$ in Eq.~\eqref{eq:Oqq}, which are built from three collinear fields along the three directions defined by the beams and the outgoing hadronic jet. 
The Wilson coefficients of the operators are determined by calculating the
$\qq \to \gamma g$ and $q g \to \gamma q$ amplitudes in SCET and in QCD. The matching calculation is greatly simplified by the fact that all of the on-shell SCET diagrams are scaleless and vanish in dimensional regularization. In the $\overline{\mathrm{MS}}$ subtraction scheme, this allows us to directly read off the result for the Wilson coefficient from the fixed order calculation in the full theory. To this end, we use the paper \cite{Arnold:1988dp} which gives the result for the virtual corrections to $\qq \to \gamma g$ and $q g \to \gamma q$ at one loop. In the effective theory, this result corresponds to the bare Wilson coefficient squared. After normalizing to the tree-level and removing divergences by renormalization, we then obtain the result for the one-loop hard function. For the annihilation channel,
the result is
\begin{multline}
H_{q \bar q}(p_T, v, \mu) = 1+\left(\frac{\alpha_s}{4\pi}\right)\left\{ -\left(2 C_F + C_A \right)\ln
^2\frac{p_T^2}{\mu^2}
+( 4C_F \ln( v \vb)+ 6 C_F ) \ln\frac{p_T^2}{\mu^2} \right.\\
+ \frac{-336+65 \pi^2}{18} - \frac{17}{3} \ln v \ln\vb+\frac{1}{6}\ln^2(v \vb)- \frac{11}{3} \ln(v\vb) \\
\left.+\frac{(-3+2v) \ln^2\vb + (48v-26)\ln\vb + (22-48v )\ln v + (-1-2v)\ln^2 v}{6(v^2+\vb^2)} \right\} \label{hardfo}\,.
\end{multline}
The Casimirs in the second and third lines have been set to $C_A=3$ and $C_F=4/3$ for simplicity.
The result for the Compton channel is presented in Appendix \ref{app:fo}. 

To perform resummation, we need the anomalous dimension of the corresponding SCET operator. This anomalous dimension is linear in $\ln \mu$, which is characteristic for problems involving Sudakov double logarithms. For NNLL accuracy, we need the logarithmic part of the anomalous dimension to three loops and the remainder to two-loop order. The anomalous dimension of a general leading-power SCET operator for an $n$-jet process involving massless partons was given in~\cite{Becher:2009cu,Becher:2009qa} and the result has been generalized to the massive case in~\cite{Becher:2009kw,Ferroglia:2009ep,Ferroglia:2009ii}. The anomalous dimensions of SCET operators are related to infrared singularities of QCD amplitudes \cite{Becher:2009cu}. A two-loop formula for these divergences was proposed by Catani~\cite{Catani:1998bh}. However, he did not have a result for the $1/\varepsilon$ pieces at two loops. His formula was later derived in~\cite{Sterman:2002qn} and the missing piece was related to a soft anomalous dimension which was calculated to two loops in~\cite{Aybat:2006wq,Aybat:2006mz}. Recently, it was realized that there are strong constraints on the infrared divergences, in particular from soft-collinear factorization and collinear limits of amplitudes~\cite{Becher:2009cu,Gardi:2009qi, Becher:2009qa,Dixon:2009ur}. These constraints explain the two-loop result for the soft anomalous dimension obtained earlier \cite{Aybat:2006wq,Aybat:2006mz} and in our case completely determine the anomalous dimension to three loops.

We need the result for the three-jet operators of the form $\bar{\chi}_{1} \, {\cal A}^\nu_{2\perp} \chi_3$, as in Eq.~(\ref{eq:Oqq}). In the color-space formalism  \cite{Catani:1996jh,Catani:1996vz} used in these papers, the RG  equation has the form
\begin{align}\label{anomdim}
\frac{\rd}{\rd\ln \mu}\, | \widetilde C(\{ p_{\bar{q}} ,p_{q},p_g \}, \mu) \rangle 
&= \bm{\Gamma}(\{\underline{p}\},\mu) | \widetilde C(\{ p_{\bar{q}} ,p_{q},p_g \}, \mu) \rangle \\
&= \left [ \sum_{i\neq j}\,\frac{\bm{T}_i\cdot\bm{T}_j}{2}\,
\gamma_{\rm cusp}(\alpha_s)\,\ln\frac{\mu^2}{-s_{ij}} 
+ \sum_{i}\,\gamma^i(\alpha_s) \right]  | \widetilde C(\{ p_{\bar{q}} ,p_{q},p_g \}, \mu) \rangle \,, \nonumber
\end{align}
where $s_{ij}\equiv 2\sigma_{ij}\,p_i\cdot p_j+i0$, and the sign factor $\sigma_{ij}=+1$ if the momenta $p_i^\mu$ and $p_j^\mu$ are both incoming or outgoing, and $\sigma_{ij}=-1$ otherwise. The Wilson coefficients only depend on the large components of the momentum, so 
$p_i^\mu \to \frac{1}{2}( {\red \bar{n}_i} \cdt p_{i}) \,{\red n_i^\mu}$, where ${\red n_i}={\red n_1}$, $\red n_2$ or $\red n_J$ is the light-like reference vector in the direction of the appropriate parton. The color-generators are $(\bm{T}_q^a)_{\alpha\beta}=t^a_{\alpha\beta}$, $(\bm{T}_{\bar{q}}^a)_{\alpha\beta}=-t_{\beta\alpha}^a$, $(T_g^a)_{bc}=-i f^{abc}$. The anomalous dimension coefficients entering the above equation were given to three loops in \cite{Becher:2009qa}. The single-parton terms involving $\gamma^i$ depend only on the representation of the $i$th parton and are given by two anomalous dimensions $\gamma^q$ and $\gamma^g$. Note that these anomalous dimensions are different from $\gamma^{f_q}$ and $\gamma^{f_g}$, which are relevant for the evolution of the PDFs near the end-point (see Eq.~(\ref{eq:AP}) below).

The above form Eq.~(\ref{anomdim}) is exact at least up to three-loop
order. Terms involving the conformal ratios introduced in \cite{Gardi:2009qi} can only appear for four or more partons and an additional constant term is ruled out by considering constraints from collinear limits \cite{Becher:2009qa}. Furthermore,
for the operators we consider, there is only a single color
structure: the three fields are contracted with $t_{i j}^a$, where
$i,j$ and $a$ are the colors of the anti-quark, quark, and gluon
fields respectively in the operator. The Wilson coefficient in color space can thus be written in the form
\begin{equation}
| \widetilde C(\{ p_{\bar{q}} ,p_{q},p_g \}, \mu) \rangle = t_{ij}^a  \widetilde C(\{ p_{\bar{q}} ,p_{q},p_g \}, \mu)\,.
\end{equation}
Plugging in the explicit form of the generators, the RG-equation becomes
\begin{multline}
\frac{\rd}{\rd\ln \mu} \widetilde C(\{ p_{\bar{q}} ,p_{q},p_g \}, \mu)
= \left\{ \gamma_{\rm cusp}(\alpha_s)\,\left [ -  \frac{C_A}{2} \, \left(\ln\frac{\mu^2}{-s_{{\bar{q}}g}}+ \ln\frac{\mu^2}{-s_{qg}} \right)
\right. \right.\\ \left.\left.
-\left(C_F-\frac{C_A}{2}\right) \, \ln\frac{\mu^2}{-s_{\qq}}\right] + 2\gamma^q + \gamma^g \right\} \, \widetilde C(\{ p_{\bar{q}} ,p_{q},p_g \}, \mu)\,.
\end{multline}

The Wilson coefficients ${\widetilde C}$ depend on the directions ${\red n_i^\mu}$ as well as on the momenta $p_i^\mu$. 
However, the dependence only arises via the large momentum components,  $p_i^\mu \to \frac{1}{2} (\nb_{\red i}\cdt p_i) {\red n_i^\mu}$.
At leading power, products of these large components are equal to the usual Mandelstam invariants.  That the hard function only depends on these invariants is also clear since it arises from a  calculation entirely within the full theory, which has no access to the light-cone reference vectors, so that we know that the final answer can only depend on Lorentz-invariant
products of the momenta. Moreover, there is only one dimensionless ratio at threshold, 
so we know that $H$ can only depend on $v=1+\hat{t}/\hat{s}$. From the above result for the RG equation of the Wilson coefficient, we then find that the hard function for the $\qq$ satisfies 
\begin{equation}
\frac{\rd H_{\qq} (p_T, v, \mu)}{\rd \ln\mu} = \left[ (2 C_F + C_A)
\gamma_{\mathrm{cusp}} \ln\frac{p_T^2}{\mu^2} - 2 C_F
\gamma_{\mathrm{cusp}} \ln(v \vb)+ 2 \gamma^H -\frac{\beta(\alpha_s)}{\alpha_s}  \right] H_{\qq} (p_T,v, \mu)\, ,
\end{equation}
where $\gamma^H = 2\gamma^q + \gamma^g$. The extra $\beta(\alpha_s)$ piece comes from our normalization of
the hard function; it compensates for the scale dependence of the $\alpha_s$ factor in the leading order cross-section 
(see Eq.~(\ref{sigdef})).
The solution is
\begin{multline}
H_{\qq} (p_T, v, \mu) = \frac{\alpha_s(\mu_h)}{\alpha_s(\mu)}\exp \left[ \left( 4 C_F + 2 C_A \right) S (\mu_h, \mu) -
2 A_H (\mu_h, \mu) \right] \label{hev} \\
\times \left( \frac{p_T^2}{\mu_h^2} \right)^{- (2 C_F + C_A) A_{\Gamma}
(\mu_h, \mu)} \left( v \vb \right)^{2 C_F A_{\Gamma} (\mu_h, \mu)} H_{\qq} (p_T,
v, \mu_h) \,,\qquad\qquad
\end{multline}
where $H (p_T, v, \mu_h)$ has the perturbative expansion in $\alpha_s$ given in Appendix \ref{app:fo}. For the Compton channel,
\begin{multline}
H_{q g} (p_T, v, \mu) =\frac{\alpha_s(\mu_h)}{\alpha_s(\mu)} \exp \left[ \left( 4 C_F + 2 C_A \right) S (\mu_h, \mu) -
2 A_H (\mu_h, \mu) \right] \label{hevqg} \\
\times \left( \frac{p_T^2}{\mu_h^2} \right)^{- (2 C_F + C_A) A_{\Gamma}
(\mu_h, \mu)} \left( v^{2 C_A}\vb^{2C_F} \right)^{ A_{\Gamma} (\mu_h, \mu)} H_{q g} (p_T,v, \mu_h)\,.\qquad\qquad
\end{multline}

The functions $S (\nu, \mu)$ and $A (\nu, \mu)$ are the same as in previous 
papers \cite{Becher:2006mr,Becher:2008cf}, with a factor of $C_F$ factored out of the cusp anomalous dimension in
$S (\nu, \mu)$ and $A_{\Gamma} (\nu, \mu)$. That is
\begin{align}
S (\nu, \mu) &= - \int_{\alpha_s (\nu)}^{\alpha_s (\mu)} d \alpha
\frac{\gamma_{\mathrm{cusp}} (\alpha)}{\beta (\alpha)} \int_{\alpha_s
(\nu)}^{\alpha} \frac{\rd \alpha'}{\beta (\alpha')}\,, &
A_{\Gamma} (\nu, \mu) &= - \int_{\alpha_s (\nu)}^{\alpha_s (\mu)} d \alpha
\frac{\gamma_{\mathrm{cusp}} (\alpha)}{\beta (\alpha)}\,.
\end{align}
$A_H (\nu, \mu)$ is the same  as $A_{\Gamma}$ but with $\gamma^H$ replacing
$\gamma_{\mathrm{cusp}}$. Explicit expressions for these functions in RG-improved perturbation theory can be found in \cite{Becher:2006mr}.

\subsection{Soft functions \label{sec:soft}}
We consider the soft functions next. The Lagrangian of the soft sector of SCET is identical to the standard QCD Lagrangian, so the calculation of the soft matrix element is the same as in QCD. They are determined by matrix elements of time-ordered products of three Wilson lines.
Rewriting Eq.~(\ref{softdef}), for the two channels,
\begin{align}
\langle 0 |\,{\rm Tr}\, \bar{\bm T}\! \left[ (Y_1^\dagger \, Y_Jt^a\,Y_J^\dagger Y_2 )(x_-)\right ] \,
{\bm T}\! \left[( Y_2^\dagger \, Y_J t^a Y_J^\dagger Y_1) (0)\right ] | 0 \rangle \label{soft1}
&= 
 C_F N_c 
 \int_0^{\infty} \rd k_+ \, e^{-i k_+ ({\red \bar{n}_J} \cdot x)/2}\, S_{\qq}(k_+) \,, \nonumber \\
 \langle 0 |\,{\rm Tr} \, \bar{\bm T}\! \left[ (Y_1^\dagger \, Y_2t^a\,Y_2^\dagger Y_J )(x_-)\right ] \,
{\bm T}\! \left[( Y_J^\dagger \, Y_2 t^a Y_2^\dagger Y_1) (0)\right ] | 0 \rangle
&=
 C_F N_c 
 \int_0^{\infty} \rd k_+ \, e^{-i k_+ ({\red \bar{n}_J} \cdot x)/2}\, S_{q g}(k_+) \,.\non
\end{align}
The soft functions for the $\qq$ and $q g$ channels differ only by which representation of $SU(3)$ is associated with which
direction. In particular, the position  $x_-^\mu = ({\red \bar{n}_J}\cdt x )\frac{ {\red n_J^\mu}}{2}$ at which they are evaluated points in the direction of the
adjoint in the $\qq \to g \gamma$ case and a triplet (or anti-triplet) in the $q g \to q \gamma$ case.

\begin{figure}[t!]
\begin{center}
\begin{tabular}{ccccc}
\psfrag{a}[B]{$n_1$}\psfrag{b}[t]{$n_2$}\psfrag{c}{$n_J$}
\includegraphics[width=0.5\textwidth]{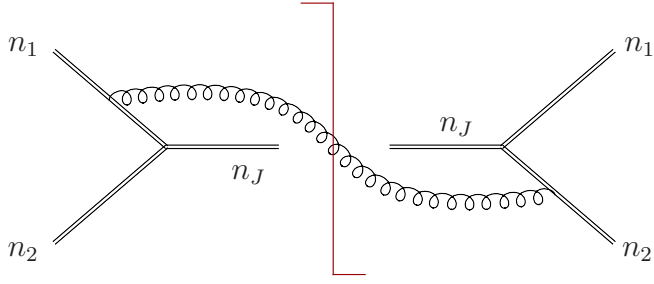}
\end{tabular}
\end{center}
\caption{Diagrams contributing to the soft function at NLO.\label{fig:softGraphs}}
\end{figure}

In dimensional regularization the virtual graphs contributing to this soft function vanish, so we are left with
real emission diagrams. These can be drawn as cuts through diagrams with a gluon being exchanged between any Wilson line at $0$ and 
any other Wilson line at $x$, as shown in Figure~\ref{fig:softGraphs}.
The soft (Eikonal) Feynman rules give a factor of
$\frac{{\red n_i^{\mu}}}{(q \cdot {\red n_i})}$ for the emission
from leg $i$, so in particular graphs involving emission and absorption into
the same leg vanish.
As indicated by the one-dimensional Fourier transforms in Eq.~\eqref{soft1}, the $x_-$ dependence means
we only need the dependence on the component of
soft radiation backward to the direction of the jet.

The non-vanishing diagrams for the $\qq \rightarrow g \gamma$ case give
\begin{multline}
S_{\qq} (k) = 2\, g^2_s \mu^{2 \varepsilon} \int \frac{\rd^{d} q}{(2\pi)^{d-1}} \delta(q^2)\theta(q_0)\delta (k - {\red n_J} \cdot q)
\\
\left. \times { \left[ \left( C_F - \frac{1}{2} C_A \right)
\frac{{\red n_1} \cdt {\red n_2}}{({\red n_1} \cdt q) ({\red n_2} \cdt q)} \right.} + {
\frac{1}{2} C_A \frac{{\red n_J} \cdt {\red n_1}}{({\red n_J} \cdt q) ({\red n_1} \cdt q)} +
\frac{1}{2} C_A \frac{{\red n_J} \cdt {\red n_2}}{({\red n_J} \cdt q) ({\red n_2} \cdt q)}} \right] \, ,
\end{multline}
and for the $q g \rightarrow q \gamma$ channel
\begin{multline}
S_{q g} (k) = 2\,g^2_s \mu^{2 \varepsilon} \int \frac{\rd^{d} q}{(2 \pi)^{d-1}} \delta(q^2)\theta(q_0)\delta(k - {\red n_J} \cdt q)\\
\left. \times { \left[ \frac{1}{2} C_A \frac{{\red n_1} \cdt {\red n_2}}{({\red n_1}  \cdt q) ({\red n_2} \cdt q)} \right.} + \left( C_F - \frac{1}{2} C_A \right)
{ \frac{ {\red n_J} \cdt {\red n_1}}{({\red n_J} \cdt q) ({\red n_1} \cdt q)} +
\frac{1}{2} C_A \frac{{\red n_J} \cdt {\red n_2}}{({\red n_J} \cdt q) ({\red n_2} \cdt q)}} \right] \, .
\end{multline}
So the calculation boils down to the evaluation of the integral
\begin{equation}
I_S(k) = \mu^{2 \varepsilon} \int \rd^{d}q\, \delta(q^2)\,\theta(q_0)
\frac{{\red n_a} \cdt {\red n_b}}{( {\red n_a} \cdt q) (
{\red n_b} \cdt q)} \delta (k - {\red n_c} \cdt q)
\end{equation}
which we need both in the case when ${\red n_c^\mu} = {\red n_a^\mu}$ and in the case when ${\red n_c^\mu}$ is different from both ${\red n_a^\mu}$ and ${\red n_b^\mu}$. 

To evaluate the integral, we write
\begin{align}
q^\mu &= 
q_+ \frac{\red n_a^\mu}{\red n_{ab}} +
q_- \frac{\red n_b^\mu}{\red n_{ab}} +
q_\perp^\mu \,,
\end{align}
with ${\red n_a} \cdt {q_\perp} = {\red n_b} \cdt {q_\perp}=0$
and 
${\red n_{i j}} \equiv {\red n_i} \cdt {\red n_j}$.
In this basis,
\begin{align}
{\red n_c^\mu} &= 
\frac{{\red n_{bc}}}{\red n_{ab}} {\red n_a^\mu}+ 
\frac{{\red n_{ac}}}{\red n_{ab}} {\red n_b^\mu}+
{\red n_{c \perp}^\mu}\,.
\end{align}
Rewriting the phase-space integration as an integral over the light-cone components and integrating over $|q_\perp|$
in $d=4-2\varepsilon$ dimensions, we find
\begin{equation}
I_S(k) = \frac{1}{2}\mu^{2 \varepsilon} 
\left( \frac{ {\red n_{ac}}{\red n_{bc}} }{ 2 {\red n_{ab}} }\right)^\varepsilon
\int \rd \Omega_{d-2}
\int\frac{\rd q_+ \rd q_-}{(q_+ q_-)^{1 + \varepsilon}} \delta (k - q_-
- q_+ + 2 \sqrt{q_+ q_-} \cos \theta )\,,
\end{equation}
where $\theta$ is the angle between ${\red \vec{n}_c^\perp}$ and $\vec{q}_\perp$. The prefactor shows that unless the three light-cone vectors are distinct the integral is scaleless and vanishes.\
Parameterizing $q_+=k y x$ and $q_-=ky {\bar x}= k y (1-x)$ and integrating over $y$ gives
\begin{equation}
I_S(k) =   \frac{1}{2} \left( \frac{ {\red n_{ac}}{\red n_{bc}} }{ 2 {\red n_{ab}} }\right)^\varepsilon 
\frac{\mu^{2 \varepsilon} }{k^{1+2\varepsilon}}
\int \rd\Omega_{d-2} 
\int_0^1 \rd x  \, x^{-1-\varepsilon} \,\bar{x}^{-1-\varepsilon} (1+2 \sqrt{x\bar{x}}\cos\theta)^{2\varepsilon}   \, . 
\end{equation} 
We can use that the integral is symmetric under $x\to \bar{x}$ to integrate only from $x=0\dots\frac{1}{2}$. Divergences then appear only in the integration region around $x=0$. After rescaling $x\to x/2$,
the integral can be expanded in $\varepsilon$, using the fact that
\begin{equation}
x^{-1-\varepsilon}  = -\frac{1}{\varepsilon} \delta(x) + \left[ \frac{1}{x}\right]_+ - \varepsilon  \left[ \frac{\ln x}{x}\right]_+  \dots \,.
\end{equation}
%
The result is
\begin{equation}
I_S(k) = \frac{2\pi^{1-\varepsilon}}{k} \left(\frac{\mu}{k}\right)^{2\varepsilon}  
\frac{1}{\Gamma(1-\epsilon)}\,\left( \frac{ {\red n_{ac}}{\red n_{bc}} }{ 2 {\red n_{ab}} }\right)^\varepsilon
\left[-\frac{1}{\varepsilon} +{\mathcal O}(\varepsilon^2)\right]\, .
\end{equation}

To expand this result in $\varepsilon$, we use
\begin{equation} \label{distexp}
\frac{1}{k} \left(\frac{\mu}{k}\right)^{2\varepsilon} = -\frac{1}{2\varepsilon}\delta(k) + \left[ \frac{1}{k} \right]_\star^{[k,\mu]}
-2\varepsilon \left[ \frac{1}{k} \ln
\frac{k}{\mu} \right]_\star^{[k,\mu]}+\cdots \,.
\end{equation}
These star-distributions are like plus distributions for a dimensionful variable~\cite{De Fazio:1999sv}. 
The superscript, as explained in~\cite{Schwartz:2007ib}, makes explicit the singular variable and the upper limit of integration:
these distributions vanish when $k$ is integrated integrated from $0$ to $\mu$.

Putting everything together, the soft functions $S_{\qq}$ and $S_{q g}$ to order $\alpha_s$ are
\begin{equation} \label{softfo}
S_{i} (k, \mu) = \delta (k) + \left( \frac{\alpha_s}{4 \pi}\right) C_i \left\{
\left[2 \ln^2\frac{2 {\red n_{12}}}{{\red n_{1 J}}{\red n_{2 J}}}-\frac{\pi^2}{3} \right]
\delta (k)
+\left[\frac{16\ln \left(\frac{k}{\mu} \sqrt{\frac{2 {\red n_{12}}}{{\red n_{1 J}} {\red n_{2 J}}}}\right)}{k}\right]_\star^{[k,\mu]} 
\right\}
\end{equation}
where the color factors for the two channels are
\begin{align}
C_{\qq}= C_F-\frac{1}{2}C_A \quad \mathrm{and}\quad   C_{qg} &= \frac{1}{2}C_A \, .
\end{align}
Note that the $\ln^2\frac{2{\red n_{12}}}{ {\red n_{1 J}} {\red n_{2 J}} }$ term
would be absent if we rewrote the star-distribution entirely in terms of
$k\sqrt{\frac{2{\red n_{12}}}{ {\red n_{1 J}} {\red n_{2 J}}}}$ rather than $k$.

To get the higher order soft function we use the constraint of RG invariance to express the soft anomalous dimension in terms of the other
anomalous dimensions (see Eq.~(\ref{softrel1}) below):
\begin{equation}
\gamma^{S_{q g}} =  \gamma^H  - \gamma^{J_q}+ \gamma^{f_g}+\gamma^{f_q}  \, .
\end{equation}
Since the three-loop hard, quark-jet, and PDF anomalous dimensions are known, this gives us $\gamma^{S_{qg}}$ to three loops.
Casimir scaling, which is known to hold up to at least three loops, then determines the other soft function anomalous dimension
to three loops as well:
\begin{equation}
\gamma^{S_{\qq}}=\frac{2C_F-C_A}{C_A} \gamma^{S_{q g}} \, .
\end{equation}

The resummation of the soft functions, from $\mu_s$ to $\mu$ follows from solving
its RG equation in Laplace space, as in~\cite{Becher:2006nr}. The result is
\begin{equation}
S_{i} (k, \mu) =
\exp[-4 C_i S (\mu_s, \mu) + 2 A_{S_{i}} (\mu_s, \mu)]
\widetilde{s}_{i} (\partial_{\blue \eta^s_i}) \frac{1}{k}
\left( \frac{k}{\mu_s} \sqrt{\frac{2 {\red n_{12}}}{{\red n_{1 J}} {\red n_{2 J}}}}   \right)^{\blue \eta^s_i}
\frac{e^{- \gamma_E {\blue \eta^s_i}}}{\Gamma ({\blue \eta^s_i})}\,, \label{eq:RGsoft}
\end{equation}
with
\begin{equation}
{\blue \eta^s_{i}} =  4 C_i A_{\Gamma}  (\mu_s, \mu)\,.\non
\end{equation}
The Laplace transform $\widetilde s(L)$ is given in Appendix \ref{app:fo}.

\subsection{Jet functions}

The quark jet function is known completely to two loops \cite{Becher:2006qw}, and its anomalous dimension to three loops \cite{Becher:2006mr}.
To order $\alpha_s$, it is
\begin{equation}
J_q (p^2, \mu) = \delta (p^2) + \left( \frac{\alpha_s}{4 \pi} \right) \left\{
\left[ C_F \left( 7 - \pi^2 \right) \right] \delta (p^2) +  \left[ \frac{4 C_F \ln\frac{p^2}{\mu^2} -
3 C_F}{p^2} \right]_{\star}^{[p^2, \mu^2]} \right\} \,.\label{jetfo}
\end{equation}
The only place where the gluon jet function has appeared previously is in the analysis of quarkonium production \cite{Fleming:2002rv,Fleming:2002sr,Fleming:2003gt}. In \cite{Fleming:2003gt}, its one-loop anomalous dimension was calculated. Here, we will compute the full order $\alpha_s$ gluon jet function and derive its anomalous dimension to order $\alpha_s^3$, although for NNLL resummation we only need the $\alpha_s^2$ result.

\begin{figure}[t!]
\begin{center}
\begin{tabular}{ccccc}
\includegraphics[width=0.25\textwidth]{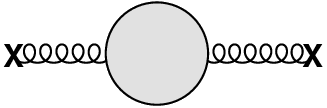} & &
\includegraphics[width=0.25\textwidth]{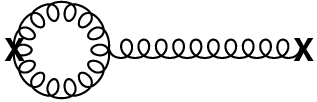} & &
\includegraphics[width=0.25\textwidth]{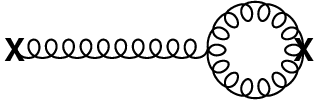} \\[0.5cm]
\includegraphics[width=0.25\textwidth]{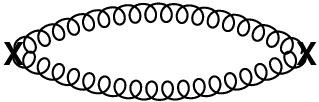} & &
\includegraphics[width=0.25\textwidth]{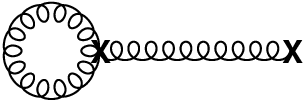} & &
\includegraphics[width=0.25\textwidth]{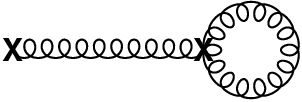} 
\end{tabular}
\end{center}
\caption{Diagrams contributing to the gluon jet function at NLO. The usual gluon self-energy contributions are represented by the first graph. In the remaining diagrams gluons are emitted from one of the Wilson lines, which are denoted by crosses.\label{fig:gluonJetGraphs}}
\end{figure}

The gluon jet function is defined by
\begin{equation}
 \langle  0 |\, {{\cal A}_{J}^a}_\perp^\mu( x) {{\cal A}_{J}^b}_\perp^\nu(0)\,   | 0 \rangle = 
(-g_\perp^{\mu\nu})\, \delta^{ab} \, g_s^2\, \int \frac{\rd^4
p}{(2\pi)^3}\,\theta(p^0)\, J_g(p^2) \,e^{-i p x}\, .
\end{equation}
The strong coupling constant $g_s$ on the right-hand side is the bare coupling; the collinear gluon fields were defined in Eq.~(\ref{eq:collfields}).
These collinear gluon operators only have non-vanishing matrix elements for intermediate collinear states. Thus, this jet function
can be thought of as the result of integrating out the collinear modes at the scale $\mu_j$. Equivalently, we can extract the jet function from the imaginary part of the time-ordered product of collinear fields
\begin{equation}
\frac{1}{\pi} {\rm Im} \left[ i \int \rd^4x\, e^{i p x}  \langle 0 | \,  {\bm T}\left\{ {{\cal A}_{J}^a}_\perp^\mu( x) {{\cal A}_{J}^b}_\perp^\nu(0) \right \} \, | 0 \rangle \right] =(-g_\perp^{\mu\nu})\, \delta^{ab}\, g_s^2\, {J}_g(p^2)\,.
\end{equation}
This second definition shows that the jet function is given by the imaginary part of the Feynman propagator in light-cone gauge, since in this gauge the Wilson lines in Eq.~(\ref{eq:collfields}) are absent.

The relevant diagrams in SCET are shown in Figure~\ref{fig:gluonJetGraphs}. In Feynman gauge all of the graphs in the bottom row vanish.
The first graph contributes to the wavefunction renormalization. Since the collinear sector of SCET is equivalent to full QCD, this graph
can be found in textbooks. In units of the tree-level result, the graph gives
\begin{equation}
I_a = \frac{\alpha_s}{4 \pi} \left(\frac{\mu^2}{-p^2}\right)^\varepsilon
\left[ \left( \frac{5}{3}  C_A - \frac{4}{3}T_F n_f \right) \frac{1}{\varepsilon
} + \frac{31}{9}C_A-\frac{20}{9} T_F n_f\right]\,.
\end{equation}
The second and third diagrams have been computed in \cite{Bauer:2001rh} and \cite{Bauer:2006mk} in Feynman gauge.  They give
\begin{equation}
I_b = I_c = \frac{\alpha_s }{4 \pi} \left(\frac{\mu^2}{-p^2}\right)^\varepsilon
C_A\left[ \frac{2}{\varepsilon^2} + \frac{1}{\varepsilon} + 2 - \frac{\pi^2}{6} \right ] \, .
\end{equation}
Adding the contributions of both diagrams and taking the imaginary part of the propagator using the relation
\begin{equation}
{\rm Im}\left[(- p^2- i\epsilon )^{-1-\varepsilon}\right]  =
-\sin(\pi \varepsilon)  (p^2)^{-1-\varepsilon}\, \, ,
\end{equation}
we obtain the bare NLO gluon jet function. The jet function can be
expanded in distributions in analogy to the soft function, see Eq.~\eqref{distexp}. Renormalizing the jet function in the $\overline{\mathrm{MS}}$ scheme, we then find
\begin{equation}
J_g (p^2, \mu) = \delta (p^2) + \left( \frac{\alpha_s}{4 \pi} \right)\left\{
\left[ C_A \left( \frac{67}{9} - \pi^2 \right) - \frac{20}{9} T_F n_f
\right] \delta (p^2) +  \left[ \frac{4
C_A \ln\frac{p^2}{\mu^2} - \beta_0}{p^2} \right]_{\star}^{[p^2, \mu^2]}\right\} \, .
\end{equation}

We have also computed all graphs in $R_\xi$ gauge and in light-cone gauge and have checked that the result for the gluon jet function is gauge invariant. In light-cone gauge, the Wilson lines $W_J(x)$ defined in Eq.~(\ref{eq:Whc}) are equal to 1 and the calculation is particularly simple, since only the first diagram contributes. The free light-cone propagator is
\begin{equation}
G_{\mu\nu}(p) = \frac{i}{p^2+i\varepsilon} \left[-g_{\mu\nu}+\frac{{\red n_\mu} p_\nu+{\red n_\mu} p_\nu}{{\red n}\cdt p+i\varepsilon} \right]\,.
\end{equation}
Note that the Mandelstam-Leibrandt (ML) prescription to regulate the ${\red n}\cdt p \to 0$ singularity is not appropriate for SCET. The ML prescription
\begin{equation}
\frac{1}{{\red n}\cdt p} \to  \frac{{\red \bar{n}}\cdt p }{{\red n}\cdt p\,{\red \bar{n}}\cdt p+i\varepsilon}
\end{equation} 
cures the collinear singularity in the propagator, but in our case this singularity has a physical meaning. The Wilson line and the associated light-cone propagators arise from expanding QCD diagrams around the large-energy limit and the choice of the $+i\epsilon$ prescription is dictated by the QCD diagrams. The loop integrals contributing to the jet function are unambiguously defined in dimensional regularization. They depend on a single four momentum $p^\mu$. If a given loop integral involves $m$ light-cone propagators it scales as $({\red n}\cdot p)^{-m}$ so that the final result of the integral is independent of the sign of the $i\varepsilon$ prescription adopted for the light-cone propagator.

To get the gluon anomalous dimension to higher order, we note that RG invariance for the $\qq \to \gamma g$ channel
implies (see Eq.~(\ref{softrel2}) below)
\begin{equation}
\gamma^{J_g}=2 \gamma^{f_q}  -  \gamma^S_{\qq} +  \gamma^H \,.
\end{equation}
Since the PDF, hard, and soft anomalous dimensions are known to three loops, this fixes 
the gluon jet anomalous dimension to three loops as well. The result is given in Appendix \ref{app:fo}.

The RG equations for the jet functions from $\mu_j$ to $\mu$ are solved with Laplace techniques~\cite{Becher:2006nr}. The results are
\begin{align} \label{eq:RGjet}
J_q (p^2, \mu) &= \exp [- 4 C_F S (\mu_j, \mu) + 2 A_{J_q} (\mu_j, \mu)]
\widetilde{j}_q (\partial_{\blue \eta_{j_q}}) \frac{1}{p^2} \left(
\frac{p^2}{\mu_j^2} \right)^{\blue \eta_{j_q}} \frac{e^{- \gamma_E {\blue \eta_{j_q}}}}{\Gamma ({\blue \eta_{j_q}})} \,, \\
J_g (p^2, \mu) &= \exp [- 4 C_A S (\mu_j, \mu) + 2 A_{J_g} (\mu_j, \mu)]
\widetilde{j}_g (\partial_{\blue \eta_{j_g}}) \frac{1}{p^2} \left(
\frac{p^2}{\mu_j^2} \right)^{\blue \eta_{j_g}} \frac{e^{- \gamma_E {\blue \eta_{j_g}}}}{\Gamma ({\blue \eta_{j_g}})}\,, \nonumber
\end{align}
where
\begin{equation}
{\blue \eta_{j_q}} = 2 C_F A_{\Gamma} (\mu_j, \mu)
\quad \mathrm{and} \quad
{\blue \eta_{j_g}} = 2 C_A A_{\Gamma} (\mu_j, \mu)  \non \, .
\end{equation}
The Laplace transforms $\widetilde{j} (L)$ are given in Appendix~\ref{app:fo}.

\subsection{Final resummed distribution in SCET \label{sec:final}}
With the hard, jet and soft functions in hand, we can now combine them
together to form the differential inclusive photon distribution.
Using the factorization formula, Eqs.~\eqref{factthm1}-\eqref{factthm2} with
the notation of Eq.~\eqref{csec}, the partonic cross section in SCET takes the
form of Eq.~\eqref{sigmafact}:
\begin{equation} \label{factHJS}
\sighat =w \widetilde\sigma(v)\, H (p_T, v ,\mu) \int \rd k J (\Mxt - (2 E_J) k,\mu) S (k,{\red n_i},\mu) \, .
\end{equation}
We will now perform this convolution.

First, note the soft function can depend, in general, on any dimensionless ratio of dot products of the
directions $\red{n_i}$. However, in Section \ref{sec:soft}, we saw that in
the formula for the resummed soft functions, Eq.~(\ref{eq:RGsoft}),
the only combination which appeared is $\frac{2 {\red n_{12}}}{ {\red n_{1 J}}{\red n_{2 J}} }$.
This ratio can be written in the suggestive
form
\begin{equation} \label{nijeq}
\frac{2 {\red n_{12}}}{{\red n_{1 J}} 
{\red n_{2 J}}} = \frac{(2 E_J)^2 \hat{s}}{\hat{t} \hat{u}}=  \frac{(2 E_J)^2}{p_T^2}\,.
\end{equation}
Since the soft function is, to all orders,  $\frac{1}{k}$ times a function of
$k \sqrt{ \frac{2 {\red n_{12}}}{{\red n_{1 J}}{\red n_{2 J}}}}$,
if we rescale $k\to \frac{k}{E_J}$ 
in Eq.~(\ref{factHJS}), all of the $E_J$ and $\red{n_i}$ dependence completely disappears, as it must.

The exact solutions of the RG equations for the soft and jet functions given in Eqs.~\eqref{eq:RGsoft} and \eqref{eq:RGjet} involve derivatives acting on a kernel which is just the relevant scale raised to a power. Because of this simple form, the convolution
in Eq.~\eqref{factHJS} can be performed analytically. In the annihilation channel, the result is
\begin{align} \label{qqform}
\sighat &=w \hat{\sigma}_{\qq}(v)
\exp \left[ \left( 4 C_F + 2 C_A \right) S (\mu_h, \mu) - 4 C_A S (\mu_j, \mu) + \left( 2 C_A - 4 C_F \right) S (\mu_s, \mu  ) \right] \nonumber \\
&\times \exp \left[ -2 A_H(\mu_h,\mu) + 2 A_{J_g}(\mu_j, \mu) + 2 A_{S_\qq}(\mu_s,\mu) \right] \\
&\times (v \vb)^{2 C_FA_{\Gamma} (\mu_h, \mu)} \left(
\frac{p_T^2}{\mu_h^2} \right)^{- (2 C_F + C_A) A_{\Gamma} (\mu_h, \mu)}
\left( \frac{\mu_j^2}{p_T \mu_s} \right)^{(4C_F - 2 C_A) A_{\Gamma} (\mu_s, \mu)}  \nonumber \\
&\times
\frac{\alpha_s(\mu_h)}{\alpha_s(\mu)}
H_{\qq} (p_T, v, \mu_h) \widetilde{j}_g (\partial_{{\red }
{\red {\blue \eta_{\qq}}}}, \mu_j) \widetilde{s}_{q q} (\ln \frac{\mu_j^2}{p_T \mu_s} + \partial_{{\blue \eta_{\qq}}},
\mu_s) \frac{1}{\Mxt} \left( \frac{\Mxt}{\mu_j^2} \right)^{\blue \eta_{\qq}} \frac{e^{- \gamma_E {\blue \eta_{\qq}}}}
{\Gamma ( {\blue \eta_{\qq}})} \, ,\nonumber
\end{align}
where
\begin{equation}
{\blue \eta_{\qq}} ={\blue \eta^j_g} +{\blue \eta^s_{\qq}}  = 2 C_A A_{\Gamma} (\mu_j, \mu) + (4 C_F - 2 C_A)
A_{\Gamma} (\mu_s, \mu)\,. \non
\end{equation}
For the Compton channel,
\begin{align} \label{qgform}
\sighat &=w \hat{\sigma}_{q g}(v)
\exp \left[ \left( 4 C_F + 2 C_A \right) S (\mu_h, \mu) - 4 C_F S (\mu_j, \mu) - 2 C_A S (\mu_s, \mu) \right] \non\\
&\times \exp \left[ -2 A_H(\mu_h,\mu) + 2 A_{J_q}(\mu_j, \mu) + 2 A_{S_{qg}}(\mu_s,\mu) \right] \\
&\times (v^{2 C_A} \vb^{2 C_F})^{A_{\Gamma} (\mu_h, \mu)} \left(
\frac{p_T^2}{\mu_h^2} \right)^{- (2 C_F + C_A) A_{\Gamma} (\mu_h, \mu)}
\left( \frac{\mu_j^2}{p_T \mu_s} \right)^{2 C_A A_{\Gamma} (\mu_s, \mu)}  \non\\
&\times
\frac{\alpha_s(\mu_h)}{\alpha_s(\mu)}
H_{q g} (p_T, v, \mu_h) \widetilde{j}_q (\partial_{\blue \eta_{q g}}, \mu_j) \widetilde{s}_{q g} (\ln \frac{\mu_j^2}{p_T \mu_s} 
+ \partial_{{\blue \eta_{q g}}}, \mu_s) \frac{1}{\Mxt} \left( \frac{\Mxt}{\mu_j^2} \right)^{\blue \eta_{q g}}
\frac{e^{- \gamma_E {\blue \eta_{q g}}}}{\Gamma ( {\blue  \eta_{q g}})} \, ,\nonumber
\end{align}
where
\begin{equation}
{\blue \eta_{q g}} = {\blue \eta^j_q}+ {\blue \eta^s_{q g}}  = 2 C_F
A_{\Gamma} (\mu_j, \mu) + 2 C_A A_{\Gamma} (\mu_s, \mu)\,. \non
\end{equation}
With these closed form expressions, it is straightforward to evaluate the differential cross section numerically.

\section{Cross-checks \label{sec:checks}}
Next, we perform some
non-trivial cross checks on the factorization theorem. First, we show that in
the threshold limit, the expression is RG invariant, that is, independent of
$\mu$. Then we will show that the singular parts of the fixed order expansion
agree with the exact NLO differential distribution. We also generate all the
plus distribution terms in $1 - w$ to NNLO, and compare to previous results.

In traditional approaches, one has access to only the renormalization
scale $\mu_R$, where $\alpha_s$ is evaluated, and the factorization
scale, $\mu_f$, where the PDFs
are evaluated. In the SCET approach, there are four scales: the hard matching scale $\mu_h$, where
the hard modes of QCD are integrated out, the jet scale,
$\mu_j$, where the collinear modes are integrated out, the soft scale $\mu_s$,
where the soft modes are integrated out, and the factorization scale $\mu_f$,
where the PDFs are evaluated.
In contrast to the scale $\mu_R$, the scales coming from the effective field theory calculation 
are all guaranteed to have natural values, since the calculation has been factorized into a series of single-scale problems.
When combining the RG evolved hard, jet, and soft functions with the PDFs, 
one evolves all of them to a common reference scale $\mu$. 

To check RG invariance, we evaluate the factorization theorem for a fixed scale $\mu_h=\mu_j=\mu_s=\mu_f=\mu$ and then show that the cross section is $\mu$-independent. To do so, we must expand around the physical, observable {\it machine} threshold,
$\MXt \rightarrow 0$, not the partonic threshold $\Mx \to0$, since
$\Mx$ gets integrated over. Recalling Eq.~\eqref{MXmx}, the invariant mass of the hadronic final state is
\begin{equation}\nonumber
\MXt= \Mxt + \frac{p_T^2}{v \bar{v}} \big[ (1-\x_1)v + (1-\x_2)\vb \big] +\cdots
\end{equation}
near the threshold. Since the SCET operators do not mix, the contributions of different partonic channels are separately RG invariant.
For the annihilation channel, we have
\begin{multline}
\frac{\rd^2 \sigma_{q \bar q}}{\rd \MXt \rd y} \propto \int \rd \x_1 \int \rd \x_2 \int \rd m^2 \int \rd k\, \alpha_s(\mu) H_{q\bar q}(p_T, v , \mu)
f_{q/N_1} (\x_1, \mu)  f_{{\bar{q}}/N_2} (\x_2, \mu) \\ \times J_g  (m^2, \mu) S_{q\bar q} (k, \mu) 
\, \delta\! \left[\MXt - \left( m^2 + (2 E_J)k  + \frac{p_T^2}{\vb}(1-\x_1) + \frac{p_T^2}{v}(1-\x_2)\right) \right]\,.
\end{multline}
We write the right-hand side in terms of $v$, which is implicitly a function of
$y, x_1, x_2$ and $\Mx$ to make contact with the partonic expressions in the previous section.
In this form, we can read off that $\MXt \rightarrow 0$ enforces $m^2
\rightarrow 0, k \rightarrow 0, \x_1 \rightarrow 1$ and $\x_2 \rightarrow 1$ so
all the various objects approach singular limits.

Since all the objects are convoluted together in this simplified factorization
theorem, it makes sense to check RG invariance in Laplace space, which turns
the convolution into a product. We define the Laplace transformed cross section as
\begin{equation}
\frac{\rd^2 \widetilde\sigma}{\rd Q^2 \rd y} = \int_0^\infty \rd\MXt \, 
\exp \left(-\frac{\MXt}{Q^2 e^{\gamma_E}}\right) \frac{\rd^2\sigma}{\rd \MXt \rd y}\, .\end{equation}
Absorbing $e^{\gamma_E}$ into this definition avoids a proliferation of  $\gamma_E$'s in the Laplace transformed expressions. 
The Laplace transforms of the soft and jet functions are defined analogously (see Eqs.~(\ref{jetlap}) and (\ref{softlap}) in Appendix 
\ref{app:fo}).

%

For the RG evolution of the PDFs, we can use
simplified Altarelli-Parisi equations near the endpoint. For the quark PDF,
\begin{align}\label{eq:AP}
\frac{\rd f_{q/N} (\x, \mu)}{\rd \ln\mu} 
& =  2 \gamma^{f_q}   f_{q/N} (\x, \mu)
+2 C_F \gamma_{\mathrm{cusp}} (\alpha) \int_x^{1} \rd x'  \frac{ f_{q/N} (x', \mu)-  f_{q/N} (\x, \mu)}{x' - \x} \, .
\end{align}
The Laplace transform
\begin{equation}
\widetilde{f}_{q/N} (\tau, \mu) = \int_0^1 \rd \x \exp\left(-\frac{1-\x}{\tau e^{\gamma_E}}\right) f_{q/N} (\x, \mu)
\end{equation}
then satisfies
\begin{equation}
\frac{\rd \widetilde{f}_{q/N} (\tau, \mu)}{\rd \ln  \mu} =
\left[2 C_F \gamma_{\mathrm{cusp}} \ln\left( \tau \right) + 2\gamma^{f_q} \right] \widetilde{f}_{q/N} (\tau, \mu)\, .
\end{equation}
The gluon PDF equation is the same with $C_F \to C_A$ and $\gamma^{f_q} \to \gamma^{f_g}$.
The Laplace transforms of the gluon jet function satisfies
\begin{align}
\frac{\rd}{\rd \ln\mu}  \widetilde{j}_g(Q^2,\mu)
&= \left[ - 2 C_A \gamma_{\mathrm{cusp}} \ln\left( \frac{Q^2}{\mu^2} \right) -2 \gamma^{J_g} \right] \widetilde{j}_g (Q^2,\mu) \,.
\end{align}
The quark jet function is the same with $C_A$ replaced by $C_F$ and $\gamma^{J_g}$ replaced by $\gamma^{J_q}$. The $\qq$ soft function satisfies
\begin{align}
\frac{\rd}{\rd \ln  \mu}  \widetilde{s}_{\qq}(\kappa, {\red n_{ij}}, \mu)
&= \left[(-4 C_F + 2C_A) 
\gamma_{\mathrm{cusp}} \ln\left( \frac{\kappa}{\mu}\,
\sqrt{\frac{2{\red n_{12}}}{{\red n_{1 J}}  {\red n_{2 J}}}} \right) - 2 \gamma^{S_{\qq}} \right]
\widetilde{s}_{\qq}(\kappa, {\red n_{ij}}, \mu)\, ,
\end{align}
where ${\red n_{ij}}\equiv {\red n_i} \cdt {\red n_j}$ as before. The other soft function, $S_{q g}$, is the same with $-4C_F + 2C_A$ replaced
by $-2C_A$ and $\gamma^{S_\qq}$ replaced by $\gamma^{S_{q g}}$.
Finally, the hard function satisfies,
\begin{align}\label{hardrg}
\frac{\rd H_{\qq} (p_T, v, \mu)}{\rd \ln  \mu} &= \left[ (2 C_F + C_A)
\gamma_{\mathrm{cusp}} \ln\frac{p_T^2}{\mu^2}-
\gamma_{\mathrm{cusp}} \ln(v^{2 C_F} \vb^{2 C_F}) + 2 \gamma^H 
-\frac{\beta(\alpha_s)}{\alpha_s}
\right] H_{\qq} (p_T, v, \mu) \,. 
\end{align}
The function $H_{q g} (p_T, v, \mu)$ is related to $H_{\qq} (p_T, v, \mu)$ by crossing. Its anomalous dimension can be obtained from the above equation by replacing $v^{2 C_F}\to v^{2 C_A}$.

Putting everything together, RG invariance requires
\begin{equation}\label{eq:RGinv}
\frac{\rd}{\rd \ln \mu} \left[
\alpha_s(\mu)\, H_{\qq} (p_T, v, \mu) 
\widetilde{j}_g (Q^2,\mu)
\widetilde{s}_{\qq} \left(\frac{Q^2}{2 E_J},{\red n_{ij}}, \mu \right)
\widetilde{f}_{q/N_1} \left( \frac{Q^2}{p_T^2} \vb, \mu \right)
\widetilde{f}_{\bar{q}/N_2} \left( \frac{Q^2}{p_T^2} v , \mu \right)
\right] = 0\, .
\end{equation}
The factor of $\alpha_s(\mu)$ is part of $\widetilde\sigma_{\qq}(v)$ and its scale-dependence cancels against the $\beta(\alpha_s)$-term in (\ref{hardrg}).
This equation imposes several constraints on the terms proportional to the universal cusp anomalous dimension $\gamma_{\mathrm cusp}$,
since the $\mu$, $p_T$, $E_J$, ${\red n_{ij}}$ and $v$-dependence must
all vanish. 
Using the above equations and the relationship between ${\red n_{ij}}$, $E_J$ and $p_T$ in Eq.~\eqref{nijeq}, we find that these constraints all hold.
For the non-cusp pieces of the anomalous dimension, Eq.~\eqref{eq:RGinv} implies
\begin{equation} \label{softrel2}
2 \gamma^{f_q} -  \gamma^{J_g} - \gamma^{S_{\qq}} +  \gamma^H = 0\,.
\end{equation}
The corresponding relation in the Compton channel reads
\begin{equation} \label{softrel1}
\gamma^{f_q} + \gamma^{f_g} -  \gamma^{J_q} -  \gamma^{S_{q g}} +   \gamma^H = 0\,.
\end{equation}
We have calculated all these anomalous dimensions to order $\alpha_s$,
verifying these relations.

Another way to check RG invariance is order-by-order in $\alpha_s$ (cf. Ref. \cite{Schwartz:2007ib} for a
similar example).
First by expanding the resummed hard, jet and soft functions,
from Eqs.~(\ref{hev}), (\ref{eq:RGjet}), and (\ref{eq:RGsoft}) 
it is easy to  check that the
matching scales $\mu_h, \mu_j$ and $\mu_s$ cancel between the fixed order
expansions, such as $H(p_T,v,\mu_h)$ and the evolution kernels, such as the
$\exp(S(\mu_h,\mu))$ factors in Eq.~(\ref{hev}). 
To combine the ingredients
together to check the overall $\mu$ dependence at order $\alpha_s$, we would also
need a perturbative expression for the PDFs, which is impossible since they 
are non-perturbative. However,
since RG
invariance should hold with any PDFs, a simple trick is
to use toy-model PDFs  with a convenient form. For example, we can define the quark
PDF  so that $f_{q/N} (\x, \mu_0) = \delta (1 - \x)$ to all orders at the scale
$\mu_0$. Then, to order $\alpha_s$, this PDF at the scale $\mu$ is
\begin{equation}
f_q (\x, \mu) = \delta (1 - \x) + \left( \frac{\alpha_s}{4 \pi} \right)
\left[ -3 C_F \ln\frac{\mu_0^2}{\mu^2} \right] \delta (1 - \x) + \left(
\frac{\alpha_s}{4 \pi} \right) \left[ \frac{- 4 C_F \ln \frac{\mu^2_0}{\mu^2}}{1 - \x} \right]_+ \,.
\end{equation}
The gluon PDF is the same with $3C_F\to \beta_0$ and $4C_F \to 4C_A$.
Then one can convolute these simple PDFs
together with the NLO hard, jet, and soft functions to verify $\mu$ independence to order $\alpha_s$.

Finally, the fixed order expressions for hard, jet, and soft functions when combined at the scales $\mu_h=\mu_j=\mu_s=p_T$,
should reproduce all the singular terms in the exact parton level NLO amplitudes from the full standard model. 
The results for these singular terms are presented in Appendix \ref{app:nlo}
and agree precisely with~\cite{Gordon:1993qc}. In addition, working to order $\alpha_s^2$, we can derive all of the 
terms singular in $1-w$ at NNLO. Previous results with NLL resummation~\cite{Kidonakis:1999hq}
were able to predict only some of these singularities.

\begin{figure}[t]
\begin{center}
\includegraphics[width=\textwidth]{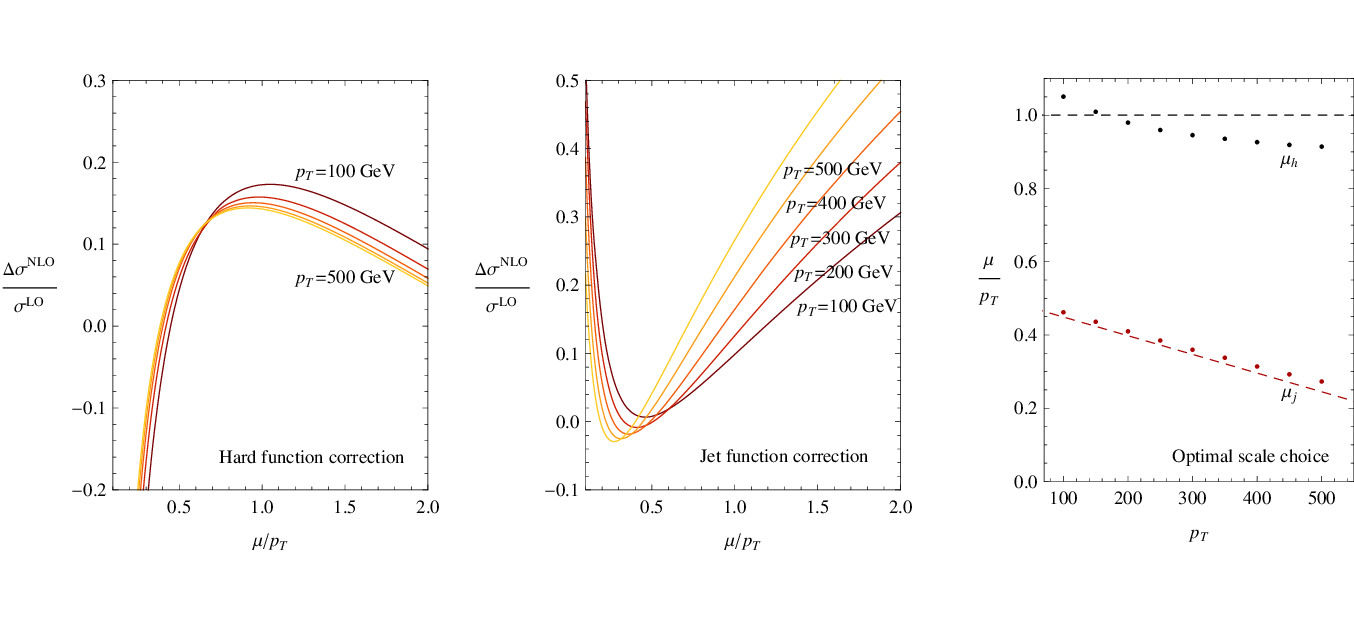}
\end{center}
\vspace*{-1.4cm}  
\caption{Size of the hard and jet function one-loop corrections as a function of the scale for different values of $p_T$ at $\ecm =$1960 GeV. The right panel shows the optimal scale choice at different $p_T$, with the dashed lines denoting our default choice, Eq.~(\ref{mujchoice}).}
\label{fig:jetFOvary}
\end{figure}

\section{Scale choices and matching \label{sec:scalechoice}}

While the resummed result is formally independent of the scales $\mu_h$, $\mu_j$, and $\mu_s$, there is residual higher-order dependence on these scales if the perturbative expansions of the hard, jet and soft functions are truncated at a finite order. 
To get a well behaved expansion, we want to evaluate each contribution at its natural scale, where it does not involve large perturbative logarithms. In a fixed order calculation, the presence of several scales can preclude such a choice, but since the hard jet and soft functions each only depend on a single scale, we are guaranteed that there are scale choices for which large logarithms are absent. 

\begin{figure}[t]
\begin{center}
\includegraphics[width=\textwidth]{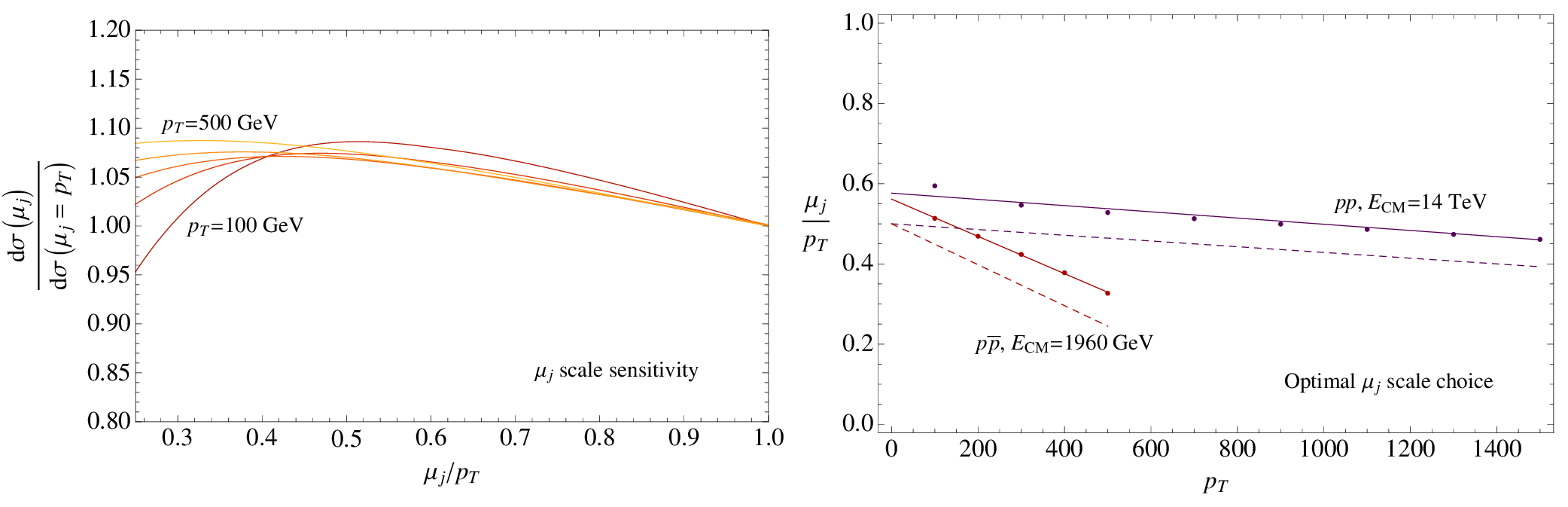}
\end{center}
\vspace*{-0.7cm}  
\caption{Determination of $\mu_j$.
On the left is the relative cross section for
variations of $\mu_j$ around $\mu_j=p_T$ for $\ecm$=1960 GeV. The other scales are chosen to be $\mu_h=\mu_f=p_T$ 
and $\mu_s = \mu_j^2/\mu_h$. 
On the right, the values of $\mu_j$ which minimize the scale variation at various $p_T$'s are shown for the Tevatron and the  LHC. The solid lines show a linear regression to
the points, and the dashed line is our default choice, Eq.~\eqref{mujchoice}.}
\label{fig:jetvary}
\end{figure}

By examining the form of the resummed distribution, Eqs. \eqref{qqform} and \eqref{qgform},
it can be seen that the hard, jet and soft scales appear in the cross section only through the combinations
\begin{equation}
\frac{p_T^2}{\mu_h^2},\quad
\frac{\Mxt}{\mu_j^2} , \quad
\frac{\Mxt}{p_T \mu_s}\,.
\end{equation}
Picking $\mu_h = p_T$, $\mu_j = \Mx$ and $\mu_s = \Mxt/p_T$ as the canonical
scales would guarantee the absence of large logarithms, but this choice is problematic. To see the problem, recall that $\Mxt = \frac{1}{w} \frac{p_T^2}{\vb} (1 - w)$, and the parton-level distribution  is singular at $w=1$. This singularity is integrated over since the hadronic final states are integrated over,
and the final photon $p_T$ spectrum is completely regular. Near  $w\sim 1$, the mass of the partonic final state $\Mx$ becomes small and with the choice $\mu_j=\Mx$ the coupling constants  $\alpha_s(\mu_j)$ and   $\alpha_s(\mu_s)$ are evaluated at arbitrarily low scales. Because of the Landau pole singularity of the running coupling the convolution integrals are then no longer well-defined. 
The $w\sim 1$ part of the integrand is suppressed by the resummation, and the contribution from this region of the integral should only amount to a power-suppressed correction to the overall result. However, the spurious power corrections arising in the integration can be of a lower order (and thus of larger size) than the physical power corrections to the factorization theorem~\cite{Beneke:1995pq}.

In \cite{Catani:1996yz} it was argued that these spurious singularities are particularly strong in momentum space and that it is therefore preferable to perform resummation 
in moment space. However, the effective theory framework allows us to completely avoid the need to evaluate the coupling at unphysically small scales. It is not necessary to eliminate the logarithms in the partonic cross section, what matters is that the final physical cross section is free of large logarithms. Instead of choosing the jet scale $\mu_j$ at the integrand level we should choose the scale after the convolution with the PDFs. That is, instead of setting $\mu_j = \Mx$, 
the appropriate jet scale is something like the average mass of a jet contributing to the cross section. 

To get a sense of what an appropriate average jet scale should be, let us consider some limits.
At very large $p_T$, the relevant scale in the physical cross section is the mass of the hadronic final state, so the choice $\mu_j^2 \sim \MX^2 = \ecm^2 (1-p_T/p_T^{\rm max})$ is appropriate. However, at moderate $p_T$, which is relevant in practice, the appropriate scale choice is less clear. In this case, the partonic mass $\Mx$ at a given $p_T$ value can vary kinematically over a large range, $0<\Mx<\MX$, but the fall-off of the PDFs near $x\to 1$ suppresses the region of large $\MX$ and hence of
large $\Mx$ as well. Consequently,
the partonic threshold region of small $\Mx$ is enhanced.
This dynamical enhancement of was pointed out by \cite{Appell:1988ie, Catani:1998tm} and was studied in detail \cite{Becher:2007ty} for the case of Drell-Yan production. It was found that this enhancement is mostly effective for relatively high Drell-Yan masses, which corresponds to high $p_T$ in our case.

\begin{figure}[t]
\begin{center}
\includegraphics[width=0.50\textwidth]{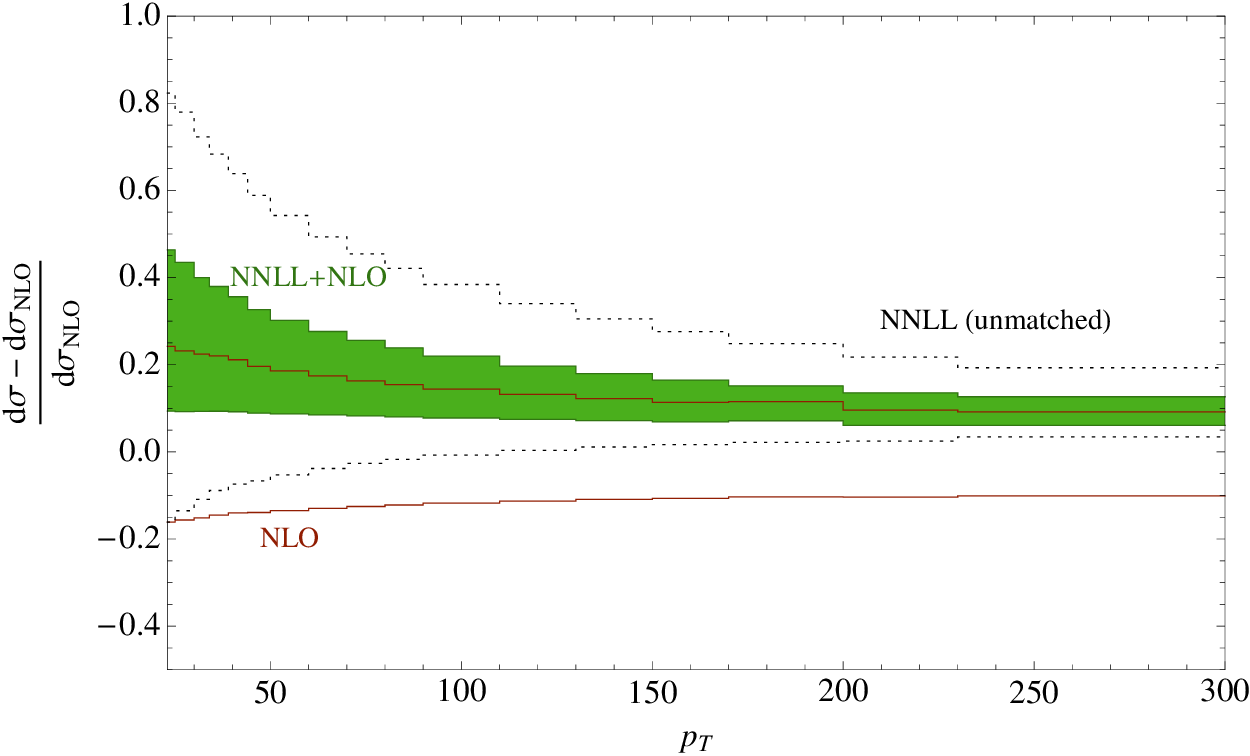} 
\end{center}
\vspace*{-0.7cm}  
\caption{Reduction of the factorization scale dependence through matching. The dotted lines show the $\mu_f$
scale uncertainty of the unmatched 
NNLL result, the red lines show the NLO uncertainty, and the green band shows the $\mu_f$ uncertainty on NNLL matched to NLO.
This is for $p\bar p$ collisions at $\ecm=1960$ GeV integrated over $-0.9<y<0.9$.}
\label{fig:matchplot}
\end{figure}

Since we cannot perform the convolution integrals analytically, we
will determine the  appropriate choice of $\mu_j$ numerically,
following two different procedures. On the one hand, we can study the
size of the corrections which arise at the different scales. Once the
scale is chosen appropriately, no large logarithms and associated
large corrections should arise. To study the size of the corrections,
we take the factorized cross section, Eq.~\eqref{sigmafact}, as a function of $\mu$, integrate over the partonic phase space, and compare the tree-level value to the result obtained after including the one-loop corrections to either the hard, jet, or soft function. The result is shown in Figure \ref{fig:jetFOvary}. The figure shows that the hard corrections are moderate if they are evaluated at $\mu_h\sim p_T$, as expected. The jet function corrections are small at a lower value. Looking at the middle panel, we find that the choice $\mu_j \sim \frac{p_T}{2}$ is reasonable for small $p_T$. For larger values of $p_T$, the the optimal scale  $\mu_j$ is lower than $\frac{p_T}{2}$. To be concrete, let us define the optimal scale as the scale which minimizes (or in the case of the hard function maximizes)  the correction. The right-hand panel shows that the choices
\begin{align}
\mu_h &= p_T \,,  \nonumber \\
\mu_j & = \frac{p_T}{2}\left (1-2 \frac{p_T}{\ecm}\right)  \,, \label{mujchoice} 
\end{align}
provide a good approximation to the optimal scale choice as a function of $p_T$. For the soft scale, we choose $\mu_s= \mu_j^2/\mu_h$ as our default choice and we have checked that the corrections are moderate for this scale choice. The plots in Figure \ref{fig:jetFOvary} are for the Tevatron case, but we have also checked that the above scale choices are also valid at the LHC, and that the optimal scales for the $\qq$ and $q g$ channels are compatible.

The reasoning behind the above procedure for choosing the scale is that there are no large logarithms and thus no large corrections if the scale is chosen appropriately. Another criterion for a good scale choice is that the residual scale dependence should be small. To explore this, we set $\mu_h=\mu_f=p_T$ and $\mu_s=\mu_j^2/\mu_h$ so that the cross section only depends on the single scale $\mu_j$. We then choose $\mu_j$ such that the distribution is minimally sensitive to variations in $\mu_j$ away from its canonical value. In the first panel of Figure \ref{fig:jetvary} we show the photon $p_T$ spectrum integrated over $|y|<1$ at the Tevatron for various values of $\mu_j$. For simplicity, we normalize to the cross section at $\mu_j=p_T$, but since we are only interested in the scale dependence, the normalization is arbitrary. 
The position of the maxima fit nicely along the curve $\mu_j = 0.56( p_T - 1.6 \frac{p_T}{\ecm})$. The same procedure at
the LHC (14 TeV) gives a best fit $\mu_j = 0.57( p_T  - 1.9 \frac{p_T}{\ecm})$. In the right panel, we show these points, the fits, and our simple scale choice, Eq.~\eqref{mujchoice}. It is comforting that also this criterion leads to similar results.

\begin{figure}[t]
\begin{center}
\includegraphics[width=\textwidth]{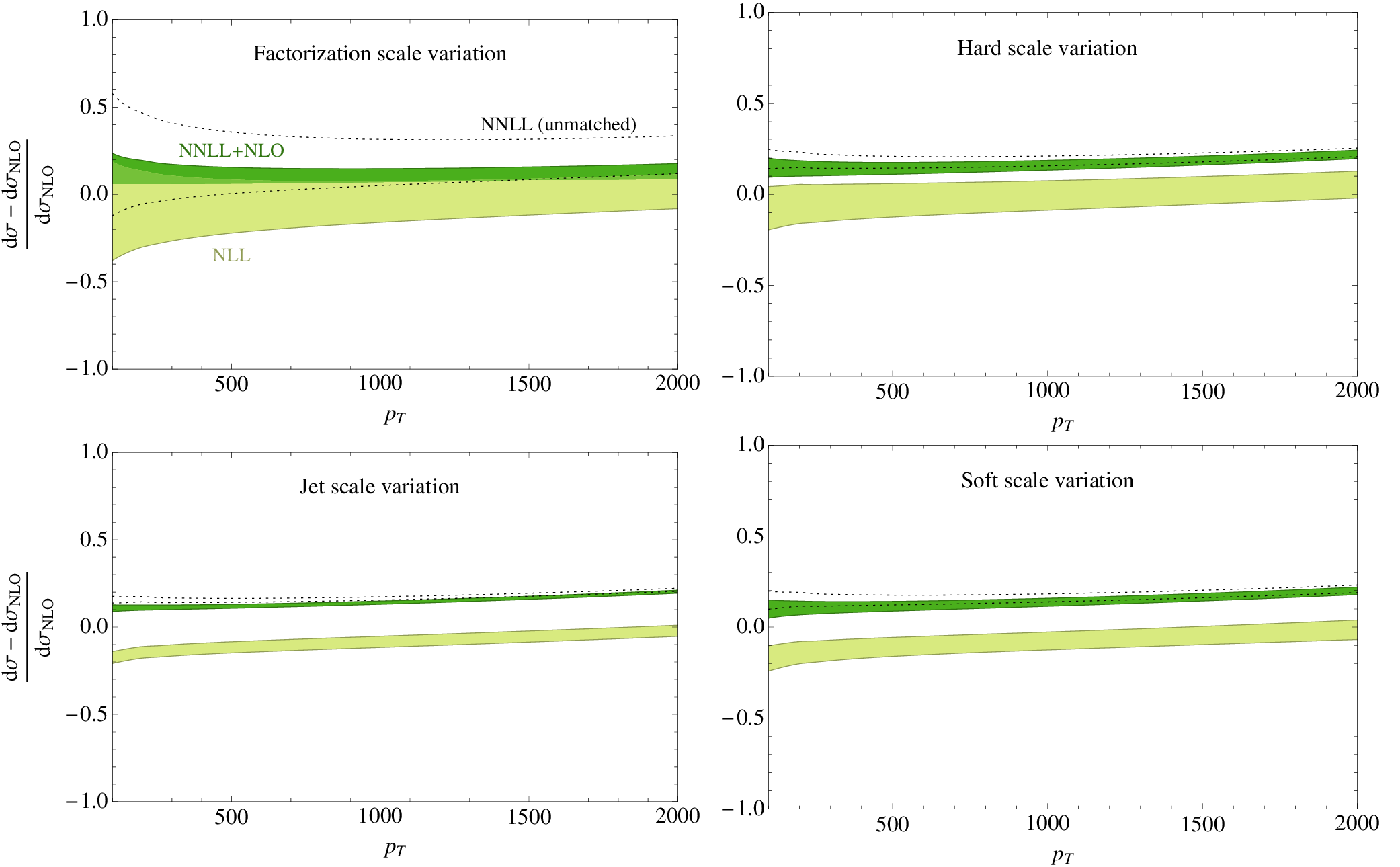} 
\end{center}
\vspace*{-0.7cm}  
\caption{Scale variations at the LHC (14 TeV). The lighter bands are NLL and the darker bands are NNLL matched to NLO.
The unmatched NNLL curves are shown as dotted lines.}
\label{fig:bandplot}
\end{figure}

So that the results from SCET agree with the NLO partonic cross section in the appropriate limit,
power corrections must be added through matching. Because of the peculiar kinematics
of the threshold limit, this must be done with some care.
The factorization theorem in SCET is derived in the limit where the momentum fractions $x_1$ and $x_2$ of the incoming partons,
and the partonic threshold variable $w$, are all close to 1. The
resummed cross section is therefore only formally $\mu_f$ independent for very large $p_T$,
in contrast to the fixed-order cross section, which has additional terms to cancel the $\mu_f$ dependence exactly, but only
works to order $\alpha_s$.
These additional terms are
not singular in the threshold variables
and therefore
not reproduced by the leading-power factorization theorem. In the phenomenologically relevant regime,
$x_1,x_2$ and $w$ may not be close to $1$, and the residual scale dependence might not be small.
This NLO part of the $\mu_f$ sensitivity can be removed as we match to the NLO partonic cross section, if the factorization scale
in the NLO cross section is varied appropriately. For the matching, we use
\begin{equation}\label{match}
\left(\frac{\rd^2\sigma}{\rd v \rd w}\right)^{\rm matched} = \left(\frac{\rd^2\sigma}{\rd v \rd w}\right)^{\rm NNLL} - \left(\frac{\rd^2\sigma}{\rd v \rd w}\right)^{\rm NNLL}_{\mu_h=\mu_j=\mu_s=\mu_f}  +  \left(\frac{\rd^2\sigma}{\rd v \rd w}\right)^{\rm NLO}_{\mu_f}\,.
\end{equation}
The subscripts of the last two terms  mean set all
scales equal to the relevant value of $\mu_f$.
Having $\mu_f$ in the matching terms vary in this way
significantly reduces the overall $\mu_f$ dependence, as can be seen in
Figure \ref{fig:matchplot}. This figure also shows that the factorization scale uncertainty at large $p_T$ is smaller than the uncertainty on the NLO cross section, even without matching.

With the canonical scales and matching procedure established, we estimate the higher order uncertainty by varying the scales by a factor of $\frac{1}{2}$ to $2$ around their default values.
The resulting uncertainties are shown in Figure~\ref{fig:bandplot}. The overall uncertainty is dominated by the factorization scale variation. The small bands from variations of $\mu_j$ and $\mu_s$ should be taken with a grain of salt. The above discussion shows that our scale choice is close to the point with minimal scale sensitivity, so that the scale variation might underestimate the higher order corrections. Also, we observe that the one-loop corrections to the soft function happen to be small in our case, much smaller than what was found in other applications.

\section{Results \label{sec:results}}

\begin{figure}[t]
\begin{center}
\begin{tabular}{ccc}
\includegraphics[width=0.48\textwidth]{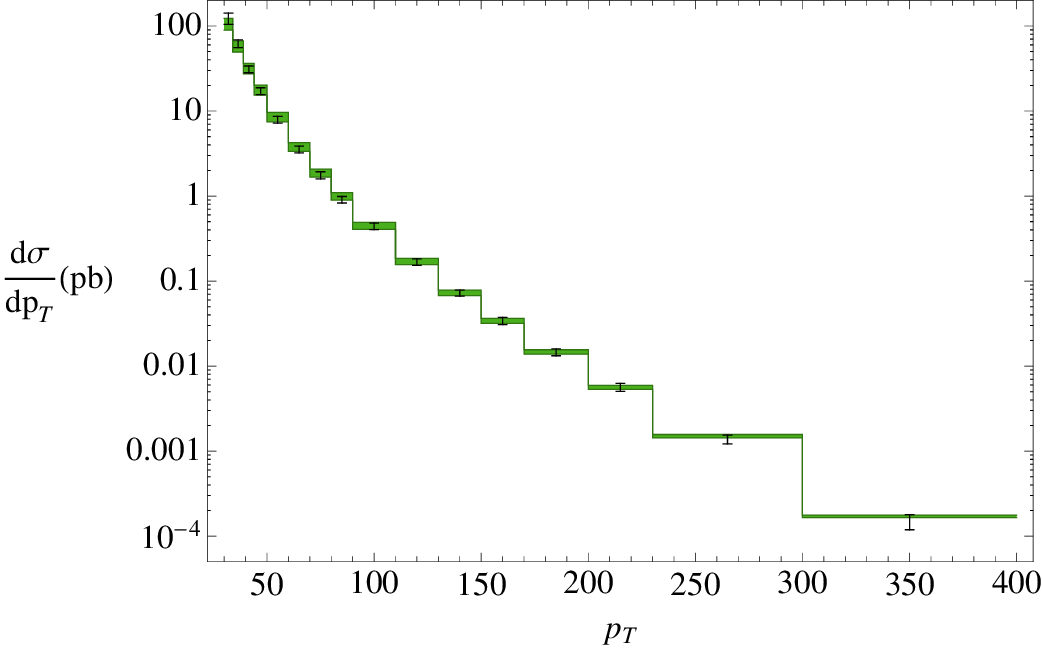}
\includegraphics[width=0.48\textwidth]{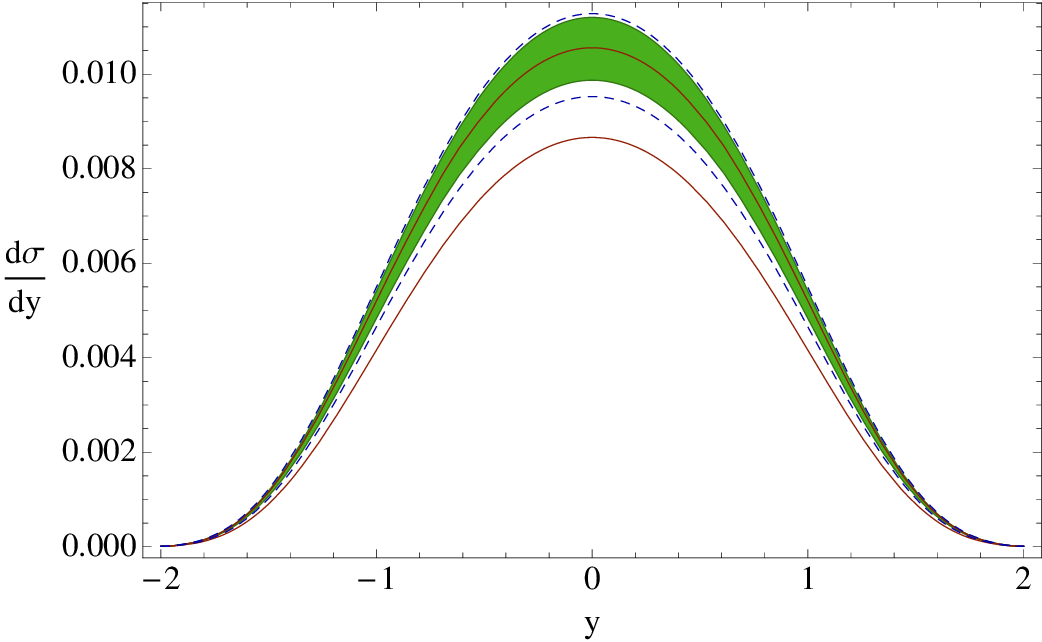}
\end{tabular}
\end{center}
\vspace*{-0.7cm}
\caption{Direct photon distributions at the Tevatron, compared to SCET. Green bands are scale uncertainty. On the left, 
comparison is made to CDF data. 
On the right, the rapidity distribution is shown for $p_T=200$ GeV. The SCET prediction, matched to NLO, is
compared to  the scale uncertainty on the NLO prediction (solid red lines) and to the PDF uncertainty (dashed blue lines).
}
\label{fig:ptandrap}
\end{figure}

To compare to data, we need to deal with the important experimental issue of
photon isolation. To account for isolation we use the Monte Carlo program {\sc jetphox}. This program
includes both the NLO partonic cross section and a fragmentation contribution,
applying a user-defined isolation criteria. To correct the SCET distributions for isolation,
fragmentation, and finite NLO effects, we match to {\sc jetphox}, {\em i.e.} we use the output of this program for the NLO cross section in the matching relation Eq.~(\ref{match}). To compare to the D0 data~\cite{Abazov:2005wc}, we attempt to match their isolation criterion by demanding less than $10\%$ of the energy
in a cone of $R=0.4$ around the photon be hadronic. For the CDF data~\cite{DelucaSilberberg:2009zz,Aaltonen:2009ty},
we require less than $2$ GeV of energy inside the $R=0.4$ cone. Some studies of sensitivity to isolation parameters can
be found in~\cite{DelucaSilberberg:2009zz} and we do not attempt to reproduce them here.

\begin{figure}[t]
\begin{center}
\includegraphics[width=\textwidth]{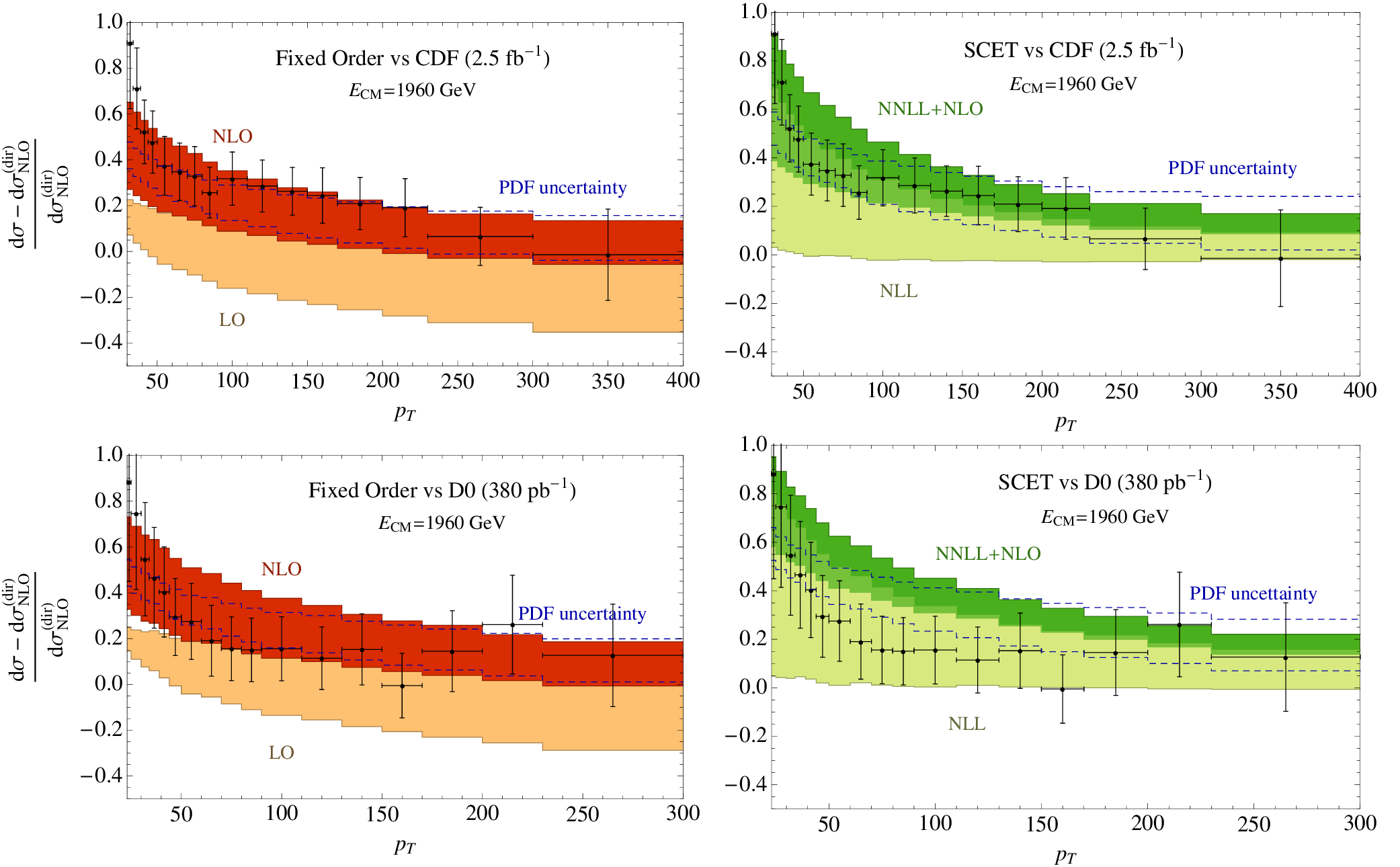} 
\end{center}
\vspace*{-0.7cm}  
\caption{Fixed order and resummed comparison to D0 and CDF data. Left plots show the LO and NLO scale uncertainties.
Right plots show the SCET predictions with NLL resummation or with NNLL
resummation matched to fixed order. 
The dashed blue lines are PDF uncertainties.
The curves are all corrected for isolation, fragmentaion, and hadronization as described in the text, while the reference distribution $\rd \sigma_\mathrm{NLO}^\mathrm{(dir)}$ is the fully inclusive NLO distribution without corrections.}
\label{fig:cdfd0}
\end{figure}

\begin{figure}[t]
\begin{center}
\includegraphics[width=\textwidth]{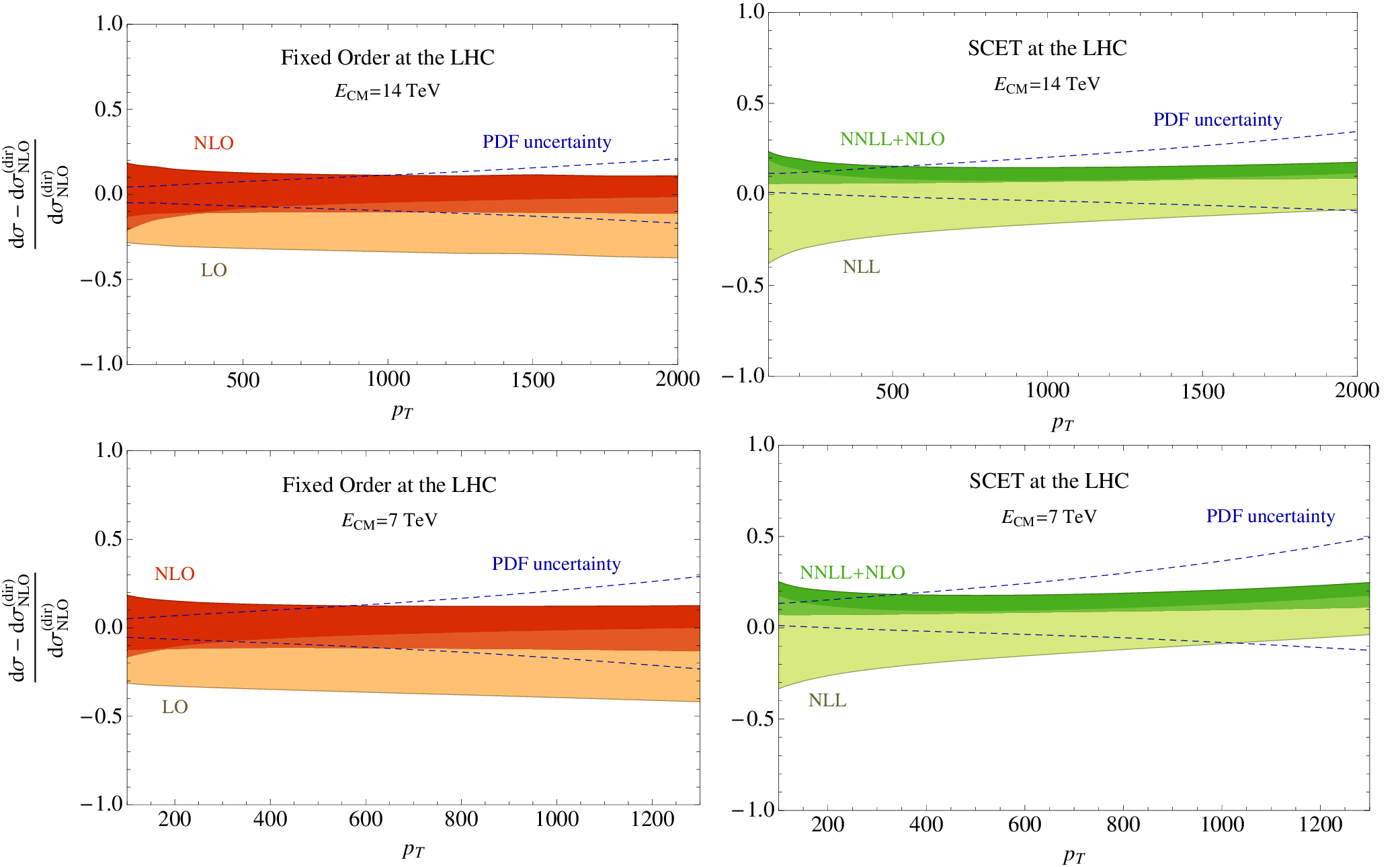}
\end{center}
\vspace*{-0.7cm}  
\caption{Predictions for the inclusive direct photon distribution at the LHC. Left plots show the LO and NLO  scale uncertainty.
Right plots show the SCET predictions with NLL resummation or with NNLL resummation matched to fixed order. 
The dashed blue lines are PDF uncertainties.
No correction for isolation or hadronization is included. In contrast to Figure~\ref{fig:cdfd0}, here NLO refers to the inclusive direct photon distribution whose central value is identical to the reference distribution $\rd \sigma_\mathrm{NLO}^\mathrm{(dir)}$.}
\label{fig:lhc}
\end{figure}

In addition, we apply to all the Tevatron theoretical calculations an overall rescaling of 0.913 (taken from~\cite{DelucaSilberberg:2009zz,Aaltonen:2009ty})
to account for underlying event, multiple interactions, and hadronization. 
The D0 data corresponds to $380$ pb${}^{-1}$ of integrated luminosity at $\ecm=1960$ GeV, integrated over $-0.9<y<0.9$.
The CDF data corresponds to 2.5 fb${}^{-1}$ of integrated luminosity at $\ecm=1960$ GeV, integrated over $-1<y<1$.
For all calculations, including {\sc jetphox} and scale uncertainties, we use the MSTW 2008 NNLO PDFs \cite{Martin:2009iq}. The rationale behind this choice is that our calculation includes the dominant NNLO corrections.

The scale uncertainties for the fixed order result include variation of the factorization scale $\mu_f$, the renormalization scale $\mu_R$, and a fragmentation scale $M'$. 
The fragmentation scale is related to collinear singularities in final state photon emission from, 
for example, $\qq$ final states, which are relevant starting at NLO. For simplicity, we call all these scales $\mu$ and vary
them together.
We define the NLO uncertainty as the maximum and minimum value of the prediction from varying these scales between 
$\frac{1}{2} p_T<\mu<2 p_T$. For the SCET prediction, we vary the jet, hard, soft and factorization scales. The largest uncertainty is from the factorization scale variation, even after the proper matching to NLO (see previous section),
and so we use the $\mu_f$ dependence for the SCET uncertainty bands. 
Again, we take the maximal and minimal values along the range $\frac{1}{2} p_T< \mu_f < 2 p_T$.

Figure \ref{fig:ptandrap} shows the $p_T$ and rapidity distributions at the Tevatron. The $p_T$ distribution is compared to CDF data~\cite{DelucaSilberberg:2009zz,Aaltonen:2009ty}
and the rapidity distribution only to the inclusive NLO result and the PDF uncertainties. 
No comparison to data has been made in the rapidity plot because all of the published
Tevatron data differential in the photon rapidity is differential in the jet rapidity as well, for which our factorization theorem does not apply. Nevertheless, such a comparison would be interesting as there is a
significant discrepancy between the SCET result and the NLO prediction.

For more detail, we show in Figure \ref{fig:cdfd0}
the normalized $p_T$ spectra and compare to CDF~\cite{DelucaSilberberg:2009zz,Aaltonen:2009ty} and D0 data~\cite{Abazov:2005wc}. In this figure and in the LHC plots in Figure \ref{fig:lhc}, we normalize to $\sigma^{\rm NLO}$, the inclusive NLO direct photon cross section, without isolation cuts and fragmentation contributions, evaluated with the default scale choices. The left plots
show the LO and NLO distributions, matched to {\sc jetphox}, with the blue dashed lines indicating NLO PDF uncertainties (from the MSTW 2008 NNLO PDFs). The right plots show the
predictions from SCET at NLL and NNLL, also matched to {\sc jetphox}, with the appropriate PDF uncertainties included as well.
Note that at high $p_T$, the scale uncertainty for the SCET result is smaller than the PDF uncertainty, while for the {\sc jetphox}, NLO result, it is not.

Figure \ref{fig:lhc} shows the SCET prediction at the LHC, with $\ecm= 14$~TeV and $\ecm=7$~TeV, integrated over $-1 < y < 1$.
The two energies give results that are qualitatively very similar, but of course, the higher energy machine
would be capable of producing higher $p_T$ photons.
Note the large PDF uncertainties at high $p_T$. These distributions
indicate that the quark, antiquark, and gluon PDFs at large $x$
can be measured effectively with direct photon production at the LHC. The PDF uncertainties are slightly larger with the SCET cross
section than with the NLO cross section, due to slightly different scales and $x$-values where the  PDFs are evaluated.

\section{Conclusions}
We have shown how to resum the direct photon distribution using Soft-Collinear
Effective Theory. This is the first physical process calculated in SCET involving more
than two collinear directions. The factorization theorem we derived passes a number
of non-trivial checks.  
In particular, renormalization scale independence arises only after a cancellation of the dependence on angles
${\red n_i} \cdt {\red n_j}$ appearing in the soft function against $\hat{s}, \hat{t}$, and $\hat{u}$ dependence in the hard function.
We have calculated all the relevant objects to one loop, showing that the dependence on the various kinematic variables is indeed of the right form for the factorization theorem to hold.

Our closed-form expression for the distribution allows for straightforward numerical integration and comparison to data.
The agreement with Tevatron data is very good, although the comparison is complicated by the issue of photon isolation. However, 
the factorization theorem holds at large $p_T$, where the isolation is less relevant.
We also show results at higher center-of-mass energy, relevant for the LHC. There, we have found that the theoretical uncertainty is much smaller than the PDF uncertainty. Thus, the resummed direct photon distribution will be an effective tool for measuring the PDFs at the LHC. In addition, there is a significant difference between the
SCET prediction at high $p_T$ at the LHC and the NLO prediction. This is not surprising as there are large logarithms in this region which SCET 
resums to all orders. In particular, our NNLL resummed result has all the singular parts of the partonic cross section at NNLO (except for the $\delta$-function part,
which is only incompletely known). Based on experience with other processes, such as Drell-Yan or Higgs production, our result is expected to provide a good approximation to the full NNLO cross section. 

Besides being of phenomenological importance, the calculations in this paper are easily generalizable to other fundamental processes at hadron colliders. The obvious
example is $W$ or $Z$ production at high $p_T$. For this case, the factorization theorem is identical. The jet and soft functions are also the same, and
the hard function is the same up to corrections finite in $m_Z$ and $m_W$. Since $W/Z$ production is free of the complication of photon isolation,
it is cleaner phenomenologically. We are currently pursuing the analysis and will present our results elsewhere.
For direct photon production, it would also be interesting to treat the photon isolation cut in the effective theory, since it is known that the corresponding perturbative expression involves large logarithms which make the fixed order calculation of this effect problematic \cite{Catani:2002ny}.

The next steps in complexity are to consider distributions which are hadronically no longer fully inclusive, and to include outgoing jets for the proton remnants. In either case, a modified factorization theorem is needed.
Including proton remnant jets is necessary to get away from the end-point region where the leading partons carry almost all of the proton momentum. To get to the phenomenologically more interesting region of moderate momentum fractions, one needs to account for the energetic partons down the beam pipe, a problem which has only recently been considered in SCET \cite{Stewart:2009yx}. Given the progress in this field over the past few years, it now becomes possible to analyze also complicated collider processes with effective field theory, in particular processes with several hadronic jets and nontrivial kinematical restrictions on the hadronic final state.

{\subsection*{Acknowledgements}}
The authors would like to thank Randall Kelley, Eric Laenen, and Lorenzo Magnea for helpful discussions,
and Guido Bell, Stefanie Marti, Xavier Tormo and Hua-Xing Zhu for pointing out typos in earlier versions. Many of the numerical computations
in this paper were performed on the Odyssey cluster supported by the FAS Research Computing Group at Harvard University.
M.D.S. is supported in part by the Department of Energy OJI program, under grant DE-FG02-91ER40654. 
The research of T.B. was supported in part by the Department of Energy under Grant DE-AC02-76CH03000.
Fermilab is operated by the Fermi Research Alliance under contract with the U.S. Department of Energy.
\vspace{0.8cm}

\begin{appendix}

\section{Fixed Order Expansions \label{app:fo}}

In our notation, all the anomalous dimensions are expanded as series in
$\frac{\alpha_s}{4 \pi}$. 

The QCD $\beta$ function is
\begin{equation}
\beta (\alpha_s) = - 2 \alpha_s \left[ \left( \frac{\alpha_s}{4 \pi} \right)
\beta_0 + \left( \frac{\alpha_s}{4 \pi} \right)^2 \beta_1 + \left(
\frac{\alpha_s}{4 \pi} \right)^3 \beta_2 + \cdots \right]\,,
\end{equation}
where
\begin{align}
\beta_0 &= \frac{11}{3} C_A - \frac{4}{3} T_F n_f  \,,\\
\beta_1 &= \frac{34}{3} C_A^2 - \frac{20}{3} C_A T_F n_f - 4 C_F T_F n_f \, , \\
\beta_2 &= \frac{2857}{54} C_A^2 + \left( 2 C_F^2 - \frac{205}{9} C_F C_A -
\frac{1415}{27} C_A^2 \right) T_F n_f + \left( \frac{44}{9} C_F +
\frac{158}{27} C_A \right) T_F^2 n_f^2 \,.
\end{align}
The four-loop coefficient $\beta_3$ is known as well \cite{vanRitbergen:1997va,Czakon:2004bu}, and can be found for example in \cite{Becher:2006mr}.

The cusp anomalous dimensions is
\begin{equation}
\gamma_{\mathrm{cusp}} (\alpha) = \left( \frac{\alpha_s}{4 \pi} \right)
\Gamma_0 + \left( \frac{\alpha_s}{4 \pi} \right)^2 \Gamma_1 + \left(
\frac{\alpha_s}{4 \pi} \right)^3 \Gamma_2 + \cdots\,,
\end{equation}
where
\begin{align}
\Gamma_0 &= 4 \,,\\ 
\Gamma_1 &= 4 \left[ C_A \left( \frac{67}{9} - \frac{\pi^2}{3} \right) -
\frac{20}{9} T_F n_f \right]\,, \\
\Gamma_2 &= 4 \left[ C_A^2 \left( \frac{245}{6} - \frac{134 \pi^2}{27} +
\frac{11 \pi^4}{45} + \frac{22}{3} \zeta_3 \right) + C_A T_F n_f \left( -
\frac{418}{27} + \frac{40 \pi^2}{27} - \frac{56}{3} \zeta_3 \right) \right. \\
&\qquad  \left. + C_F T_F n_f \left( - \frac{55}{3} + 16 \zeta_3 \right) -
\frac{16}{27} T_F^2 n_f^2 \right] \,.
\end{align}

The anomalous dimensions describing the evolution of the quark and gluon PDFs near $x=1$ are
\begin{align}
\gamma^{f_q}_0 &= 3C_F \, ,\\
\gamma^{f_q}_1 &= C_F^2\left(\frac{3}{2} - 2 \pi^2 + 24 \zeta_3\right) + C_F C_A \left(\frac{17}{6} + \frac{22\pi^2}{9} - 12 \zeta_3\right)
-C_F T_F n_f \left( \frac{2}{3} + \frac{8\pi^2}{9}\right) \, ,\\
\gamma^{f_g}_0 &= \beta_0 \, ,\\
\gamma^{f_g}_1 &= C_A^2\left(\frac{32}{3}+12 \zeta_3\right) - \frac{16}{3} C_A T_F n_f- 4 C_F T_F n_f \, ,
\end{align}
The three loop splitting functions were calculated in \cite{Moch:2004pa,Vogt:2004mw}. Explicit expressions for anomalous dimensions $\gamma^{f_q}_2$ and $\gamma^{f_g}_2$ at three loops can be found in \cite{Becher:2007ty,Ahrens:2008nc}.

The hard function can be written as
\begin{equation}
H (p_T, \mu, v) = h (\ln\frac{p_T^2}{\mu^2}, v)\,.
\end{equation}
To order $\alpha_s^2$, it is
\begin{align}
&  h \left( L, v \right) = 1 + \left( \frac{\alpha_s}{4 \pi} \right) \left\{ -
\Gamma^H_0 \frac{L^2}{2} + (\Gamma_0 L_v^H - \gamma_0^H - \beta_0) L + c_1^H(v) \right\} \\
& + \left( \frac{\alpha_s}{4 \pi} \right)^2 
\left\{ \left( \Gamma_0^H  \right)^2 \frac{L^4}{8} +  \left( 4 \beta_0  - 3 \Gamma_0 L_v^H +   3 \gamma_0^H \right) 
\Gamma_0^H \frac{L^3}{6} \right. \non\\
& \qquad + \left[ - \Gamma_1^H - c_1^H(v) \Gamma_0^H + (\beta_0 - L_v^H \Gamma_0 +   \gamma_0^H) (2 \beta_0 - L_v^H \Gamma_0 + \gamma_0^H) \right] \frac{L^2}{2}  \non \\
& \qquad \qquad+ \left.
\left[ - 2 c_1^H(v) \beta_0 - \beta_1 - c_1^H(v) \gamma_0^H - \gamma_1^H + c_1^H(v)   \Gamma_0 L_v^H + \Gamma_1 L_v^H\right] 
L + c_2^H(v) \vphantom{\frac{L^4}{8}} \right\} \non\,.
\end{align}
As described in Section \ref{sec:HJS}, the anomalous dimensions can be extracted from the general result \cite{Becher:2009qa}. Explicity,
\begin{align}
\qquad \Gamma^H &= \left( C_F + \frac{1}{2} C_A \right) \gamma_{\mathrm{cusp}} \,,\\
\qquad  \gamma_0^H &= - \beta_0 - 6 C_F\,, \\
\qquad \gamma_1^H &= \left( \frac{256}{27} - \frac{2 \pi^2}{9} \right) C_A n_f T_F +
\left( \frac{368}{27} + \frac{4 \pi^2}{3} \right) C_F n_f T_F + \left( - 3 + 4 \pi^2 - 48 \zeta_3 \right) C_F^2  \non \\
&\phantom{=}+ \left( - \frac{692}{27} +  \frac{11 \pi^2}{18} + 2 \zeta_3 \right) C_A^2
+ \left( - \frac{961}{27} -  \frac{11 \pi^2}{3} + 52 \zeta_3 \right) C_A C_F \,.\nonumber
\end{align}
The three loop anomalous dimension for the hard function is just $\gamma^H_2=2 \gamma^q_2 + \gamma^g_2$, where the anomalous dimensions $\gamma^q$ and $\gamma^g$ were defined and given to three loops in \cite{Becher:2009qa}.

The $v$-dependence is different in the two channels. It shows up in the
functions $L_v^H$ and the constants $c_j^H$. For the annihilation channel, we find
\begin{align}
L_v^{H_{\qq}} &= C_F  \ln(v \vb)\, ,\\
c_1^{H_{\qq}}(v)&= \frac{-336+65 \pi^2}{18} - \frac{17}{3} \ln v \ln\vb+\frac{1}{6}\ln
^2(v \vb) - \frac{11}{3} \ln (v\vb)\non\\
&\qquad
+\frac{(-3+2v) \ln^2\vb + (48v-26)\ln\vb + (22-48v )\ln v + (-1-2v)\ln^2 v}{6(v^2+\vb^2)} \, ,\non
\end{align}
and for the Compton channel,
\begin{align}
L_v^{H_{\qq}} &= C_A \ln(v) + C_F \ln(\vb) \, ,\\
c_1^{H_{q g}} (v) &= \frac{-336+59 \pi^2}{18} - \frac{14}{3} \ln v \ln\vb -7 \ln^2 v - \frac{1}{3} \ln^2\vb - 8 \ln(v \vb)\non\\
&+\frac{ (3-2v)\pi^2 +4 \vb\ln v(\ln v-1)+ (3-2v)\ln\vb (\ln\vb -2 \ln v + 1) + 23 \ln\vb}{3(1+\vb^2)} \non \, .
\end{align}

The solution of the RG equation for the jet function is given in terms of its Laplace transform
\begin{equation}
\widetilde{j}(Q^2,\mu)=\int_0^\infty \rd p^2 \exp\left(-\frac{p^2}{Q^2 e^{\gamma_E}}\right) J(p^2,\mu)\,.
\label{jetlap}
\end{equation}
This Laplace transform has a perturbative expansion in $\alpha_s$ which only depends on the dimensionless ration $Q^2/\mu^2$. 
Writing $\widetilde j(Q^2,\mu) =\widetilde j (L)$, where $L=\log\frac{Q^2}{\mu^2}$, the expansion becomes very similar
to that of the hard function
\begin{align}
\widetilde{j} ( L) &= 1 + \left( \frac{\alpha_s}{4 \pi} \right)
\left[ \Gamma^J_0 \frac{L^2}{2} + \gamma_0^J L + c_1^J \right] + \left(
\frac{\alpha_s}{4 \pi} \right)^2 \left[ \left( \Gamma_0^J \right)^2
\frac{L^4}{8} + \left( - \beta_0 + 3 \gamma_0^J \right) \Gamma_0^J
\frac{L^3}{6} \right. \non\\
& \left. + \left( \Gamma_1^J + (\gamma_0^J)^2 - \beta_0 \gamma_0^J + c_1^J  \Gamma_0^J \right) \frac{L^2}{2} 
+ (\gamma_1^J + \gamma_0^J c_1^J - \beta_0  c_1^J) L + c_2^J \right]\,.
\end{align}
For the quark jet
\begin{align}
\Gamma^{J_q} &= C_F \gamma_{\mathrm{cusp}} \, ,\\
\gamma_0^{J_q} &= - 3 C_F  \, , \non\\
\gamma_1^{J_q} &= C_F^2 \left( - \frac{3}{2} + 2 \pi^2 - 24 \zeta_3 \right) +
C_F C_A \left( - \frac{1769}{54} - \frac{11 \pi^2}{9} + 40 \zeta_3 \right) +
C_F T_F n_f \left( \frac{242}{27} + \frac{4 \pi^2}{9} \right) \, ,\non\\
c^{J_q}_1 &= C_F \left( 7 - \frac{2 \pi^2}{3} \right) \non \, .
\end{align}
The three loop anomalous dimension $\gamma_2^{J_q}$ and the two loop constant
$c_2^{J_q}$ are also known \cite{Becher:2006qw}, but are not necessary for NNLL resummation.
For the gluon jet
\begin{align}
\Gamma^{J_g} &= C_A \gamma_{\mathrm{cusp}} \,,\\
\gamma_0^{J_g} &= - \beta_0 \, ,\non\\
\gamma_1^{J_g} &= C_A^2 \left( - \frac{1096}{27} + \frac{11 \pi^2}{9} + 16  \zeta_3 \right) + C_A n_f T_F \left( \frac{368}{27} - \frac{4 \pi^2}{9}  \right) + 4 C_F T_F n_f \,,
\qquad \qquad \qquad \quad \non \\
c^{J_g}_1 &= C_A \left( \frac{67}{9} - \frac{2\pi^2}{3} \right) -  \frac{20}{9} T_F n_f  \,. \non 
\non 
\end{align}
Since it is a new result, we also give the three-loop gluon-jet anomalous dimension:
\begin{align}
\gamma_2^{J_g} &= \left( - \frac{331153}{1458} + \frac{6217 \pi^2}{243} + 260
\zeta_3 - \frac{583 \pi^4}{270} - \frac{64 \pi^2 \zeta_3}{9} - 112 \zeta_5
\right) C_A^3  \\
&  + \left( \frac{42557}{729} - \frac{2612}{243} - \frac{16 \zeta_3}{27} +
\frac{154 \pi^4}{135} \right) C_A^2 n_f T_F 
+ \left( \frac{3622}{729} + \frac{80 \pi^2}{81} - \frac{448 \zeta_3}{27}  \right) C_A n_f^2 T_F^2 \non \\
&  + \left( \frac{4145}{27} - \frac{4 \pi^2}{3} - \frac{608 \zeta_3}{9} -  \frac{16 \pi^4}{45} \right) C_A C_F n_f T_F 
- 2 C_F^2 n_f T_F - \frac{44}{9} C_F n_F^2 T_F^2 \non \, .
\end{align}

The soft function $S (k, \mu,{\red n_{ij}})$ depends in addition to the scales $k$ and $\mu$ on the
angles ${\red n_{ij}}={\red n_i\cdt n_j}$ between the Eikonal lines. However, this
dependence must be universal for the factorization theorem to hold.
The Laplace transformed soft function
\begin{equation}
\widetilde{s}(\kappa,\mu,{\red n_{ij}})=\int_0^\infty \rd k \exp\left(-\frac{k}{\kappa e^{\gamma_E}}\right) S(k,\mu,{\red n_{ij}})
\label{softlap}
\end{equation}
has a perturbative expansion in $\alpha_s$ which only depends on one dimensionless ratio. 
We can write $\widetilde{s}(\kappa,\mu,{\red n_{ij}})= \widetilde{s} (L)$ where
\begin{equation} 
L=\ln \frac{k}{\mu}
\sqrt{\frac{ 2 ({\red n_1} \cdt {\red n_2})}{ ({\red n_1} \cdt {\red n_J}) ({\red n_2} \cdt {\red n_J}) }}\,.
\end{equation}
The expansion is now similar to the hard or jet functions
\begin{align}
\widetilde{s} \left( L \right) &= 1 + \left( \frac{\alpha_s}{4 \pi} \right)
\left[ 2\Gamma^S_0 L^2 + 2\gamma_0^S L + c_1^S \right] +
\left( \frac{\alpha_s}{4 \pi} \right)^2 \left[ \left( \Gamma_0^S \right)^2
2 L^4 - \left( \beta_0 - 3 \gamma_0^S \right) \Gamma_0^S
\frac{4 L^3}{3} \right. \non \\
&  \left. + 2 \left(  \Gamma_1^S + (\gamma_0^S)^2 - \beta_0 \gamma_0^S + c_1^S
\Gamma_0^S \right) L^2 + 2 (\gamma_1^S + \gamma_0^S c_1^S -
\beta_0 c_1^S) L + c_2^S \right] \,.
\end{align}
The coefficients in the annihilation channel are
\begin{align}
\Gamma^{S_{\qq}} &= \left(C_F - \frac{C_A}{2}\right ) \gamma_{\mathrm{cusp}} \,, \\
\gamma_0^{S_{\qq}} &= 0 \,, \non \\
\gamma_1^{S_{\qq}} &= \left( C_F - \frac{1}{2} C_A \right) \left(
\left( 28 \zeta_3 - \frac{808}{27} + \frac{11 \pi^2}{9} \right) C_A + \left(
\frac{224}{27} - \frac{4 \pi^2}{9} \right) n_f T_F \right) \non \, ,\\
c^{S_{\qq}}_1 &= \left( C_F - \frac{C_A}{2} \right) \pi^2 \,. \non
\end{align}
The Compton channel is identical, but with the prefactor $C_F - \frac{1}{2} C_A$ replaced by $\frac{1}{2} C_A$. This is a consequence of Casimir scaling, which holds at least to three-loop order.

\section{NLO and NNLO singular terms \label{app:nlo}}
To obtain the singular terms, the resummed results,
Eqs.~\eqref{qqform} and \eqref{qgform},
should be expanded order-by-order in $\alpha_s$. To do so, we set all the scales equal $\mu_h =
\mu_j = \mu_s = \mu_f = \mu$. In the limit of equal scales, the
various evolution factors $S(\nu,\mu)$ and $A_\gamma(\nu,\mu)$ and the
quantities
${\blue \eta_{\bar q q}}$ and ${\blue \eta_{qg}}$ all vanish. Before
setting the scales equal, we expand the kernel using
\begin{equation}
\left(1-w\right)^{-1+{\blue \eta}} = \frac{1}{{\blue \eta}} \delta(1-w) + \sum_{n=0}^\infty \frac{{\blue \eta}^n}{n!} \left[ \frac{\ln^{n}(1-w)}{1-w} \right]_+ \,, 
\end{equation}
and perform the derivatives with respect to ${\blue \eta}$. Then we
take ${\blue \eta} \to 0$. The resulting expressions are lengthy. To save space we only quote the result for $\mu=p_T$, for which we find
\newcommand{\GJS}{\Gamma^{JS}}
\begin{multline}
\frac{\rd^2 \hat{\sigma}}{ \rd v \rd w} = {\widetilde \sigma}(v)
\Bigg[  \delta (1 - w)
+ \left( \frac{\alpha_s}{4 \pi} \right) \Bigg\{ \delta (1 -
w)\Big[ \vphantom{\frac{\pi^2}{6}} c_1^H (v) + c_1^J + c_1^S -(\gamma_0^J-\gamma_0^S)\ln \vb \\ 
+   \frac{1}{2}\left(\ln^2\vb-  \frac{\pi^2}{6}\right)\GJS_0 \Big] +
(\gamma_0^J + 2 \gamma_0^S-\GJS_0 \ln\vb) \left[ \frac{1}{1 - w} \right]_+ +
\GJS_0 \left[ \frac{\ln(1 - w)}{1 - w}  \right]_+ \Bigg\} \\
+ \left( \frac{\alpha_s}{4 \pi} \right)^2  \Bigg\{ \delta(1-w) A_2  +\bigg[  
\frac{\pi^2}{3}\beta_0 \Gamma_S^0 +( c_H^1(v)+c_S^1 + c_J^1)(\gamma_J^0 +2 \gamma_S^0) -\beta_0(2  c_S^1 +c_J^1)\\
+ \frac{\pi^2}{12}\GJS_0(\beta_0 - 3\gamma_J^0 - 6 \gamma_S^0) + 2 \gamma_S^1+\gamma_J^1+(\GJS_0)^2 \zeta_3\\
-\ln \vb \left(\GJS_1 + \GJS_0(c_H^1(v) + c_J^1 + c_S^1) - \frac{\pi^2}{4}(\GJS_0)^2 + (\gamma_J^0 + 2 \gamma_S^0)^2 -\beta_0 (\gamma_J^0 + 4 \gamma_S^0)\right)\\
-\frac{1}{2} \ln^2\vb \left( \vphantom{\frac{1}{2}} 4 \beta_0 \Gamma_S^0 + \GJS_0 (\beta_0 - 3 \gamma_J^0 - 6 \gamma_S^0)\right)
- \frac{1}{2}\ln^3\vb (\GJS_0)^2  \bigg]  \left[ \frac{1}{1 - w} \right]_+\\
+ \bigg((c_1^H(v)+ c_1^J + c_1^S)\GJS_0 +\GJS_1 - \frac{\pi^2}{4}(\GJS_0)^2+(\gamma_0^J + 2\gamma_0^S)^2 \\
-\beta_0 (\gamma_0^J +4 \gamma_0^S)+\GJS_0 (\beta_0-3\gamma_0^J - 6 \gamma_0^S) \ln\vb +4 \beta_0 \Gamma_0^S \ln\vb+ \frac{3}{2} (\GJS_0)^2 \ln^2\vb\bigg)
\left[  \frac{\ln(1 - w)}{1 - w} \right]_+ \non\\
+ \bigg( \frac{1}{2}(3 \gamma_0^J + 6 \gamma_0^S-\beta_0)\GJS_0 -2\beta_0 \Gamma_0^S - \frac{3}{2} (\GJS_0)^2\ln\vb \bigg)
\left[  \frac{\ln^2 (1 - w)}{1 - w} \right]_+  +  \frac{1}{2}( \GJS_0)^2 \left[  \frac{\ln^3 (1 - w)}{1 - w} \right]_+ \Bigg\} \Bigg]  \, ,\non 
\end{multline}
where
\begin{equation}
\GJS = \Gamma^J + 4\Gamma^S \, .
\end{equation}
This formula holds for either channel, with the appropriate ${\widetilde \sigma}(v)$ and hard, jet and soft function coefficients.
The coefficient $A_2$ is not completely known, so we do not include our
partial results. We have checked that the $\left. O (\alpha_s \right)$ results
agree with~\cite{Gordon:1993qc}
and the $\alpha_s^2 \left[ \frac{\ln^3 (1 -w)}{1 - w} \right]_+$ and $\alpha_s^2 \left[ \frac{\ln^2 (1 - w)}{1 - w}
\right]_+$ results agree with~\cite{Kidonakis:1999hq}. We find a small discrepancy\footnote{
The factor $\frac{3s+t}{s}$ in the first $\frac{C_F}{T_{qg}}$ term of their Eq.~(3.8) should read $\frac{3s+2t}{s}$.}
with these authors for the $\alpha_s^2 \left[ \frac{\ln(1 - w)}{1 - w}
\right]_+$term in the Compton channel, but otherwise we confirm their results.
The $\alpha_s^2 \left[ \frac{1}{1 - w} \right]_+$ piece was not given in~\cite{Kidonakis:1999hq} because it requires NNLL resummation.

\end{appendix}

\end{document}